\documentclass[%
superscriptaddress,  
nofootinbib,
 amsmath,amssymb,
 aps,
 prD,
 reprint,
]{revtex4-1}
\usepackage[normalem]{ulem}

\usepackage{graphicx}
\usepackage{adjustbox}
\usepackage{dcolumn}
\usepackage{bm}
\usepackage{hyperref}
\usepackage{xcolor}
\definecolor{darkgreen}{rgb}{0, 0.4, 0} 
\definecolor{midgreen}{rgb}{0.5, 0.8, 0.5}
\definecolor{darkred}{rgb}{0.5, 0, 0}
\definecolor{darkblue}{rgb}{0, 0, 0.5} 

\hypersetup{linktocpage=true, colorlinks=true, citecolor={darkblue}, linkcolor={darkred}, urlcolor={darkblue} }
\hypersetup{bookmarks=true, bookmarksnumbered=true, bookmarksopen=true, bookmarksopenlevel=1 }

\usepackage{booktabs}
\usepackage{ulem}
\usepackage{bbold}
\usepackage{comment}
\usepackage{physics}
\usepackage{url}
\usepackage{mathrsfs}
\usepackage{amsmath}
\usepackage{siunitx}
\usepackage{setspace}

\newcommand{\relnetfriction}{\Xi_\mathrm{net}}
\newcommand{\passiveres}{\omega_\mathrm{res}^\mathrm{pas}}

\usepackage{titlesec}

\def\bfseries{\fontseries \bfdefault \selectfont \boldmath}

\titleformat{\section}{\raggedright\bfseries\large}{\Roman{section}.}{1em}{}
\titlespacing\section{0pt}{10pt plus 4pt minus 2pt}{6pt plus 2pt minus 0pt} 

\titleformat{\subsection}{\raggedright\bfseries}{\Roman{section}.\Alph{subsection}}{1em}{}
\titlespacing\subsection{0pt}{10pt plus 4pt minus 2pt}{0pt plus 2pt minus 0pt} 

\makeatletter

\renewcommand{\fnum@figure}{\textbf{Figure \thefigure}}
\renewcommand{\fnum@table}{\textbf{Table \thetable}}
\makeatother

\newcommand\blfootnote[1]{%
  \begingroup
  \renewcommand\thefootnote{}\footnote{#1}%
  \addtocounter{footnote}{-1}%
  \endgroup
}

\begin{document}

\title{Hair cells in the cochlea must tune resonant modes to the edge of instability without destabilizing collective modes}

\author{Asheesh S.\ Momi}%
\affiliation{Department of Physics,  Yale University,  New Haven,  CT 06520}
\affiliation{Quantitative Biology Institute,  Yale University,  New Haven,  CT 06520}

\author{Michael C.\ Abbott}%
\affiliation{Department of Physics,  Yale University,  New Haven,  CT 06520}
\affiliation{Quantitative Biology Institute,  Yale University,  New Haven,  CT 06520}

\author{Julian Rubinfien}%
\affiliation{Department of Physics,  Yale University,  New Haven,  CT 06520}
\affiliation{Quantitative Biology Institute,  Yale University,  New Haven,  CT 06520}

\author{Benjamin B.\ Machta}%
 \email{benjamin.machta@yale.edu}
 \affiliation{Department of Physics,  Yale University,  New Haven,  CT 06520}
\affiliation{Quantitative Biology Institute,  Yale University,  New Haven,  CT 06520}

\author{Isabella R.\ Graf}%
 \email{isabella.graf@embl.de}
\affiliation{Department of Physics,  Yale University,  New Haven,  CT 06520}
\affiliation{Quantitative Biology Institute,  Yale University,  New Haven,  CT 06520}
\affiliation{Developmental Biology Unit, European Molecular Biology Laboratory, 69117 Heidelberg, Germany}

\date{19 July 2024. v2: 18 December 2024}

\begin{abstract}
\noindent 
Sound produces surface waves along the cochlea's basilar membrane.
To achieve the ear's astonishing frequency resolution and sensitivity to faint sounds, dissipation in the cochlea must be canceled via active processes in hair cells, effectively bringing the cochlea to the edge of instability.
But how can the cochlea be globally tuned to the edge of instability with only local feedback?
To address this question, we use a discretized version of a standard model of basilar membrane dynamics, but with an explicit contribution from active processes in hair cells.
Surprisingly, we find the basilar membrane supports two qualitatively distinct sets of modes: a continuum of \textit{localized} modes and a small number of collective \textit{extended} modes.
Localized modes sharply peak at their resonant position and are largely uncoupled.
As a result, they can be amplified almost independently from each other by local hair cells via feedback reminiscent of self-organized criticality.
However, this amplification can destabilize the collective extended modes; avoiding such instabilities places limits on possible molecular mechanisms for active feedback in hair cells.
Our work illuminates how and under what conditions individual hair cells can collectively create a critical cochlea.
\end{abstract}

\maketitle

\section*{Introduction}
\noindent The human cochlea is a spiral-shaped organ in the inner ear, where sound is converted into electrical signals.
The cochlea can detect sounds with frequencies across three orders of magnitude (20--20\,000Hz) and more than a trillion-fold range in power (0--130dB), down to air vibrations on the order of an angstrom.
After entering the cochlea, sound waves become surface waves along the basilar membrane (BM), depositing most incident energy in a frequency-specific location~\cite{reichenbach2014physics}.
\blfootnote{\\ To appear in PRX Life, accepted 5 December 2024.}

Dissipation in the cochlea, through friction and viscous loss,  limits frequency resolution and sensitivity. 
To counter dissipation, the cochlea contains active force-generating mechanisms~\cite{kemp1979evidence,dallos1992active,martin1999active,ren2011measurement}. 
Active processes are performed by hair cells, small sensory structures that line the BM.
For overly strong hair cell activity, the BM becomes unstable to spontaneous oscillations.
When activity almost cancels friction, the cochlea is highly sensitive to weak amplitude signals, and frequency selectivity is high.
This barely-stable regime is thus ideal for sound processing, but appears to require fine-tuning.
Here we ask how hair cells can find this operating regime.

Past models have provided insight into possible mechanisms for tuning individual hair cells~\cite{nadrowski2004active,roongthumskul2011multiple,camalet2000auditory}.
In particular, these papers studied how single hair cells can find a Hopf bifurcation, a transition between a stable and an unstable oscillatory regime.
The discussion often focuses on bullfrog hearing where there is no cochlea, and hair cells act as relatively independent mechanical oscillators~\cite{eatock1987adaptation}.
Conversely, models of the mammalian cochlea typically ignore tuning and operate under the assumption that active processes have globally reduced friction to near zero~\cite{Julicher2003}.

In this work, we argue that assuming that hair cells cancel friction for all frequencies and positions is neither necessary nor feasible, and we instead seek to understand how they can find an operating region with the desired properties of the dissipation-free state.
Friction only dominates the dynamics of the cochlea precisely at resonance because passive mechanics are underdamped \cite{lighthill1981energy}.
Furthermore, individual hair cells are small mechanical perturbations to the overall dynamics and their contribution to the non-dissipative mechanics is likely inconsequential.
We thus focus on the role of hair cell activity in reducing friction in a manner which is local in both space and frequency.  We show that, with some interesting caveats, this is sufficient to
bring the cochlea to a line of Hopf bifurcations where every frequency is nearly critical at a specific location~\cite{talmadge1993new,eguiluz2000essential}.

Towards this end we expand on an established model for the dynamics of the cochlea \cite{reichenbach2014physics,Julicher2003,talmadge1998modeling}, by explicitly adding mechanical contributions from active processes in hair cells.
We assume that hair cells detect the displacement of the BM and respond by exerting forces via a fast linear response kernel.
Each hair cell can also slowly adjust the strength of its active processes to find the global operating regime. 

Perhaps surprisingly, this model of the cochlea contains two distinct types of modes.
The first type, which we term \textit{localized} modes, strongly peak at particular resonant positions.
The second type instead have energy throughout the cochlea, reminiscent of standing waves, and we term them \textit{extended} modes.
Both types of modes are present in the passive system, and tuning activity is generally able to bring the localized  modes to the edge of instability. 
By contrast, the extended modes become unstable for many plausible forms of active processes.
For suitable forms of active processes, we further propose a mechanism reminiscent of self-organized criticality~\cite{sornette_critical_1992, sornette_mapping_1995}, which tunes the local activity strength to the edge of instability.

\begin{figure}
    \centering
    \includegraphics[width=\linewidth]{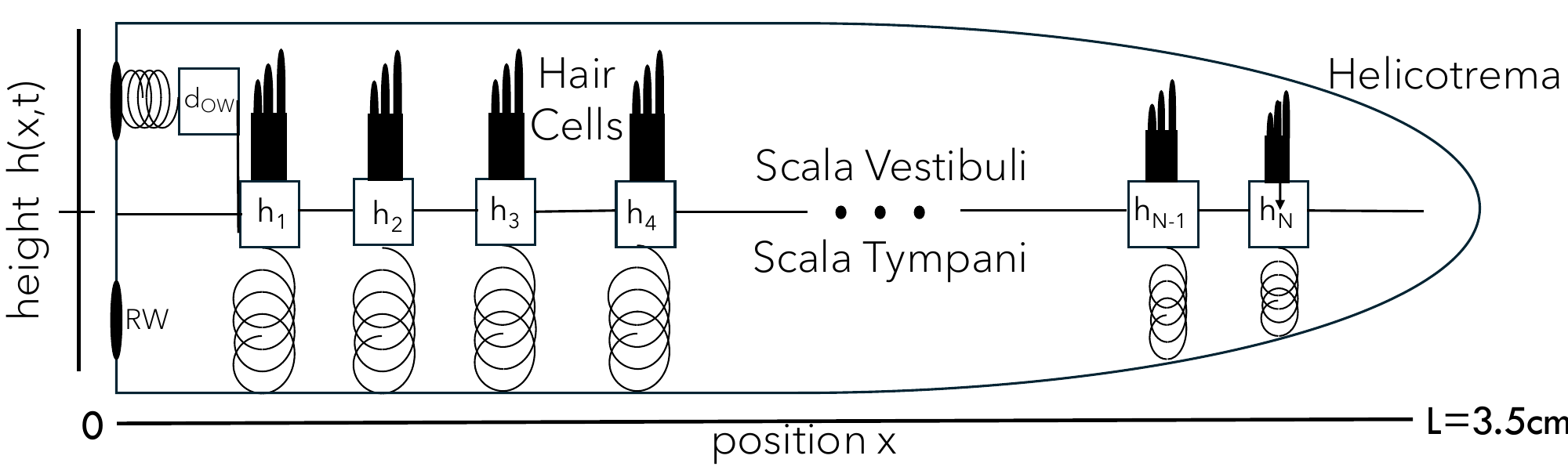}
    \caption{
    \textbf{Schematic of the model.} Above, we unroll the spiral-shaped cochlea into a model of two fluid-filled chambers partitioned by the basilar membrane (BM). Pressure waves in the fluid are accompanied by BM displacement $h(x,t)$, which we model as a set of $N$ damped harmonic oscillators, each with active driving from a hair cell. Sound input from the middle ear is via the displacement $d_\mathrm{ow}(t)$ of the oval window (OW) and a corresponding flux at the cochlea's base, which due to fluid incompressibility causes an equal and opposite flux at the round window (RW).
    The pressure difference at the helicotrema at $x=L$ is set to zero.}
    \label{fig:diagram}
\end{figure}

\section*{Results}
\noindent The passive part of our model of the cochlea has three components: the fluid which moves along the cochlea, the oval window by which the middle ear pushes this fluid, and the elastic basilar membrane which separates the cochlea into two compartments. We introduce each of these in turn and examine the resulting mode structure, before focusing on the active component, hair cells.
Our mode computation uses a discretization of the BM into $N$ components, as shown in Fig.~\ref{fig:diagram}, but to describe the fluid it is clearer to start with a continuous position $x$ along the BM.

\subsection*{Wave equation for the cochlea} 
\noindent The passive model we use is based primarily on the model from Ref.~\cite{talmadge1998modeling}, but also shares similarities with the models from Refs.~\cite{reichenbach2014physics,Julicher2003}. In line with these previous models, we approximate the cochlea as two fluid-filled compartments, the scala tympani and the scala vestibuli, which are separated by the BM (Fig.~\ref{fig:diagram}). Sound creates a fluid flux, coupled to a change in pressure  described by force balance:
\begin{equation}
\label{eq:FB}
    \rho \partial_t j(x,t)=-A_{\text{cs}} \partial_x p(x,t).
\end{equation}
Here, $j(x,t)$ is the difference in volume current between the lower and upper compartment, $p(x,t)$ is the pressure difference, $A_{\text{cs}}$ is the average cross-sectional area of a cochlear compartment, and $\rho$ is the density of water. The fluid flux propagates down the cochlea, creating a displacement of the BM, $h(x,t)$, which we call its height. The fluid flux obeys a continuity equation,
\begin{equation}
    \label{eq:cont}
2W_{\text{bm}}\partial_th(x,t)+\partial_xj(x,t)=0,
\end{equation}
where $W_{\text{bm}}$ is the width of the BM. We can eliminate $j(x,t)$ from Eq. \ref{eq:FB} and \ref{eq:cont} to arrive at a modified wave equation relating height and pressure \cite{Julicher2003,talmadge1998modeling,reichenbach2014physics}, 
 \begin{equation}
     \label{eq:wave_eq}
     \frac{2\rho W_{\text{bm}}}{A_{\text{cs}}}\partial_t^2h(x,t)=\partial_x^2p(x,t).
 \end{equation}
 
Sound enters the cochlea through the oval window, which connects to the middle ear and, in turn, the ear canal. We follow Ref. \cite{talmadge1998modeling} for the boundary conditions at the base of the cochlea ($x=0$), where the lateral displacement of the oval window $d_\mathrm{ow}(t)$ creates a flux of fluid, and eventually an equal and opposite flux in the scala tympani's round window. Via Eq.~\ref{eq:FB}, this leads to a pressure gradient:
\begin{equation}
\partial_t^2{d}_{\text{ow}}(t)=-\frac{1}{2A_{\text{ow}}}\partial_t j(0,t)=\frac{A_{\text{cs}}}{2\rho A_{\text{ow}}}\partial_x p(0,t).
\end{equation}
Here $A_\mathrm{ow}$ is the area of the oval window.
The oval window itself acts as a damped harmonic oscillator:
\begin{equation}
  \label{eq:BC}
\partial_t^2{d}_{\mathrm{ow}}(t)+\xi_{\mathrm{ow}}\partial_t{d}_{\mathrm{ow}}(t)+\omega_{\mathrm{ow}}^2d_{\mathrm{ow}}(t)=\frac{p(0,t)+G_{\mathrm{me}}P_{\mathrm{ec}}(t)}{\sigma_{\mathrm{ow}}},
\end{equation}
where  $\xi_{\text{ow}}$ is its dampening constant, $\omega_{\text{ow}}$ the middle ear resonance, $P_{\text{ec}}(t)$ the pressure in the ear canal, $G_{\text{me}}$ the gain of the middle ear, and $\sigma_{\text{ow}}$ the mass per area of the oval window. At the cochlea's apex ($x=L$), a gap in the basilar membrane (the helicotrema) suggests zero pressure difference~\cite{talmadge1998modeling}:
\begin{equation}
     p(L,t)=0.
 \end{equation}

\subsection*{Resonance from passive impedance}
\noindent To relate the height and pressure in Eq. \ref{eq:wave_eq}, we need a mechanical model of the basilar membrane and its surrounding fluid.
As is commonly assumed, we take the relationship to be local in space, where it can be written in the frequency domain via the acoustic impedance~\cite{talmadge1993new}:
\begin{equation}
    \label{eq:imped}
    \tilde{p}(x,\omega)=Z(x,\omega)\tilde{h}(x,\omega)\sigma_{\text{bm}},
\end{equation}
where $\sigma_\mathrm{bm}$ is the mass per area of the BM. Here, we separate the impedance $Z(x,\omega) = Z_{\mathrm{hc}}(x,\omega)+Z_{\mathrm{pas}}(x,\omega)$ into an active component due to hair cells, and the commonly used passive components due to stiffness, inertia, and friction:
\begin{equation}
\label{eq:Z_pas}
Z_{\mathrm{pas}}(x,\omega)=\omega_0^2e^{-2k x}-\omega^2+i\xi\omega.
\end{equation}
The stiffness decays exponentially in space \cite{emadi2004stiffness,greenwood1990cochlear}, with $\omega_0$ denoting the resonant frequency at the base of the cochlea. We deviate from the model of Ref.~\cite{talmadge1998modeling} by treating $\xi$, the dampening per unit mass, as constant for simplicity.
A tilde represents a temporal Fourier transform $\tilde{\phi}(\omega)=\int dt e^{-i\omega t} \phi(t)$.
Table \ref{tab:param} lists all the constants in this model.

The passive components of BM impedance lead to the most important feature of cochlear mechanics: spatial frequency discrimination.
Resonances occur when the two real contributions to $Z_{\mathrm{pas}}$ cancel, which happens at a position-dependent resonant frequency
\begin{equation}
\label{eq:wres}
    \passiveres(x)=\omega_0e^{-k x}.
\end{equation}
There $Z_{\mathrm{pas}}$ is purely imaginary.
If it were zero, a small change in pressure would result in an infinite change in height, and thus $Z(x,\omega)=0$ is a critical point.
However, in the real cochlea, a non-zero imaginary part due to friction limits the amplitude of near-resonant displacements. 
And while the imaginary part of the passive impedance is typically two orders of magnitude smaller than both contributions to the real part \cite{talmadge1998modeling}, it does become significant near the resonant frequency~\cite{martin2021mechanical,iwasa1997force}.
Since no passive properties can cancel it, hair cells must exert active forces to oppose friction and achieve higher sensitivity.

\begin{table}
\begin{tabular}{p{0.18\columnwidth} p{0.3\columnwidth} p{0.5\columnwidth} } 
 Symbol & Value & Description \\
 \hline\\[-2.ex]
 $\rho$ & $10^3~\si{kg\per.m^{3}}$ & Density of water \\ 
$A_{\text{cs}}$ & $1.1\times10^{-6} ~\si{m^{2}}$ & Average cross-sectional area of a cochlear partition\\ 
$A_{\text{ow}}$ & $3.2\times10^{-6} ~\si{m^{2}}$ &  Area of oval window (OW)\\ 
$W_{\text{bm}}$ & $2.9\times10^{-4}~\si{m}$ & Average width of the BM \\ 
$\xi_{\text{ow}}$ & $500~\si{s^{-1}}$ & OW dampening constant \\ 
$\xi$ & $100~\si{s^{-1}}$ & BM dampening constant  \\ 
$\omega_{\text{ow}}$ & $2\pi\times1500~\si{rad\per s}$ & OW resonant frequency \\ 
$\omega_{0}$ & $2\pi\times20800~\si{rad\per s}$ & Highest resonant frequency of the BM  \\ 
$k$ & $1.38\times10^{2}~\si{m^{-1}}$ & Stiffness decay coefficient  \\ 
$\sigma_{\text{ow}}$ & $18.5~\si{kg\per m^{2}}$ & Effective areal density of the OW  \\ 
$\sigma_{\text{bm}}$ & $5.5\times10^{-2}~\si{kg\per m^{2}}$ & Areal density of the BM  \\ 
$G_{\text{me}}$ & $21.4$ & Gain of the middle ear  \\
$L$ & $3.5\times10^{-2}~\si{m}$ & Length of the BM  \\
\end{tabular}
    \caption{\textbf{Numerical parameters of the human cochlea used in  Eq. \ref{eq:FB}-\ref{eq:Z_pas}.} All parameter values are taken from Talmadge et al., Ref.~\cite{talmadge1998modeling}. Variants of their model with additional parameters are discussed in Appendix \ref{talm}.}
    \label{tab:param}
\end{table}

\begin{figure*}
\centering
\includegraphics[width=\linewidth]{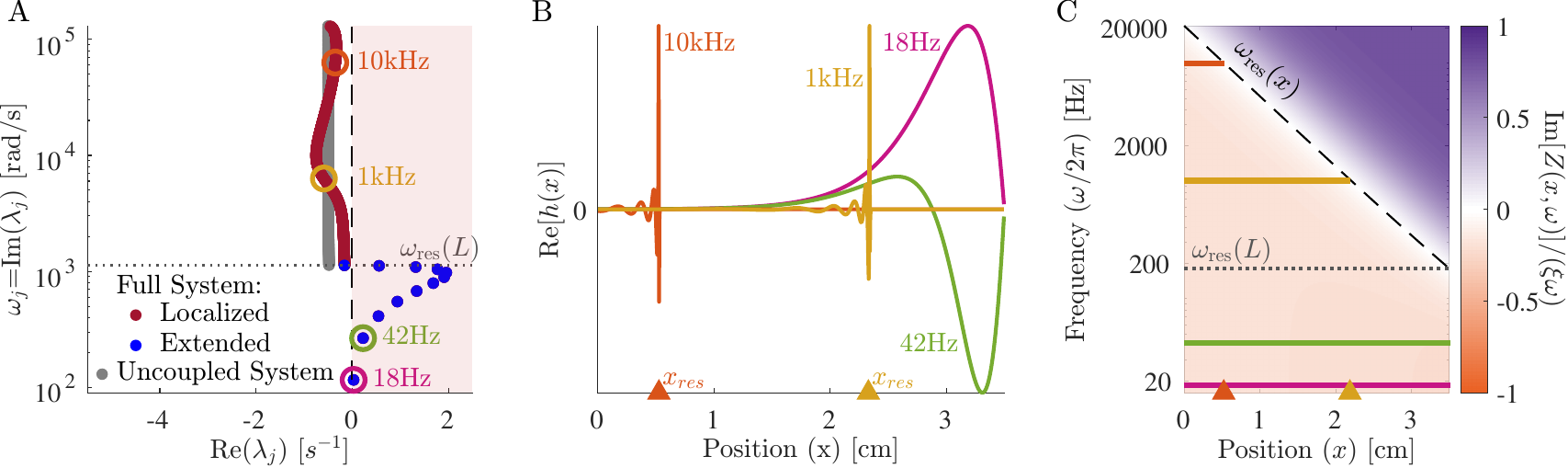}
\caption{\textbf{The cochlea exhibits a near-continuum of \textit{localized} modes plus a discrete set of \textit{extended} low-frequency modes.} (A) The eigenvalue structure of oscillating cochlear modes. An eigenvalue's imaginary part determines the oscillation frequency, and the real part determines stability. 
We define the localized modes (red) as the ones which have a resonant position within the cochlea, i.e., those for which $\lvert \Im(\lambda_j) \rvert > \omega_{\text{res}} (L)$.
Extended modes (blue) are those with $\lvert \Im(\lambda_j) \rvert <\omega_{\text{res}} (L)$, so their resonant position would be past the end of the cochlea; e.g., the $18$~Hz mode would be resonant at a position of $x=5.1$~cm.
There are $12$ extended modes, and $N-12$ localized modes, approaching a continuum of resonant frequencies at large $N$.
The uncoupled system (gray) displays the eigenvalues of $N$ independent harmonic oscillators with stiffness, mass, friction, and active force identical to each BM segment (i.e. the roots of $Z(x_n,i\lambda)=0$).
(B) Eigenvectors corresponding to the circled eigenvalues. We show the localized modes for $1$~kHz and $10$~kHz and the two lowest-frequency extended modes.
(C) Colour map of normalized $\Im[Z(x,\omega)]$, the effective net friction, across frequencies and position.
On the left of resonance (dashed black line), active processes add more energy than passive friction removes, leading to a negative effective friction (orange).
All plots have an active hair cell response kernel $g \propto e^{-r\Delta t}$ from Eq.~\ref{eq:g-exp} with $\alpha=2$, $C_{99}(x)$ from Eq.~\ref{eq:C99}, and $N=1000$. Fig.~\ref{fig:big} repeats panels A and C for other choices of response kernel $g$ defined in Eqs.~\ref{eq:g-exp}, \ref{eq:perf_adap}, and \ref{eq:IGkern}.}
    \label{fig:stab}
\end{figure*}

\subsection*{Active hair cell contributions}

\noindent  
The passive part of the model we have described so far closely follows previous models of the cochlea~\cite{talmadge1998modeling, Julicher2003, reichenbach2014physics}. To take into account active processes, a common approach is to assume that they cancel friction for all frequencies and positions, equivalent to setting $\xi=0$~\cite{Julicher2003,talmadge1998modeling}. 
Implementing this strict condition would, however, require an instantaneous derivative response, discussed below.

Here, we instead allow hair cells to respond over a finite time.
We write the active contribution from hair cells to  Eq.~\ref{eq:imped} in terms of a generic linear response kernel $g$ and a dimensionless strength $C$: 
\begin{equation}
\label{eq:imped_hc}
Z_{\mathrm{hc}}(x,\omega)=C(x)\,\tilde{g}(x,\omega).
\end{equation}
This term contributes an active pressure $p_{\mathrm{hc}}$ best understood in the time domain:
\begin{equation}
p_{\mathrm{hc}}(x,t)=\sigma_{\mathrm{bm}}C(x)\int_0^{\infty} d\Delta t\:g(x,\Delta t)\:h(x,t-\Delta t).\label{eq:p_hc}
\end{equation}
The response kernel $g(x,\Delta t)$ characterizes the active force's dependence on past displacement, and $C(x)$ controls the strength of the hair-cell force.
We initially take $g(x,\Delta t)\propto e^{-r(x)\Delta t}$, indicating that the hair cell integrates height for a time
of order $1/r(x)$. This dependence could model, for instance, the concentration of
calcium ions which enter when  hair cells are displaced and are pumped out at rate $r$, with the accumulated concentration
inside the cell controlling molecular motors~\cite{camalet2000auditory}.
To function at both
high and low frequencies, we might expect the timescale to vary
like $1/\passiveres(x)$ along the BM, and we should certainly
expect that different molecular mechanisms will be employed to respond
at 200Hz vs 20kHz.
Hair cell activity can perturb the resonant frequency away from $\passiveres(x)$, and in general we define $\omega_\mathrm{res}(x)$ by
\begin{equation}
    \label{wres}
    \Re[Z(x,\omega_\mathrm{res}(x))]=0.
\end{equation}

Adjusting the strength of hair cell activity $C(x)$ is not sufficient to cancel friction for all $x$ and $\omega$, as the frequency dependence of $Z_{\mathrm{hc}}(x,\omega)$ comes from $\tilde g(x,\omega)$ which does not in general match the linear passive term, $i\xi\omega$.
But adjusting $C(x)$ does allow us to set $Z(x, \omega_\mathrm{res}(x))=0$, cancelling friction along a line in the position-frequency plane.
With some important caveats discussed below, we will show that this is sufficient to make the cochlea highly sensitive.
To investigate the resulting properties, we assume for now that a fixed fraction $f$ (usually 99\%) of the passive friction is cancelled at the resonant frequency:
\begin{equation}
\label{eq:C99}
C_{100f}(x)=f \, \frac{-\xi\omega_\mathrm{res}(x)}{\ \Im[\tilde{g}(x,\omega_\mathrm{res}(x))]}.
\end{equation}
However, cancellation of a large fraction like $f=0.99$ requires extreme fine-tuning, as a 2\% increase in $f$ would make the system unstable everywhere.
Instead of fine-tuning $C(x)$ directly, we will later show how hair cells can use local information to robustly tune $C(x)$ to bring the cochlea to the edge of instability without fine-tuning.
But first, we will discuss the qualitative behaviour of the model at $C_{99}(x)$.

\subsection*{Mode structure of the cochlea} 
 
\noindent We took a numerical approach to better understand the features of our model, discretizing the BM into $N$ units located at $x_n = Ln/N$. 
Eqs.~\ref{eq:wave_eq},\ref{eq:BC} can then be written as a set of coupled first-order differential equations. In matrix form, these become $\partial_t{X}=\hat{J}X$, where the state vector $X$ concatenates $d_\mathrm{ow}(t)$, $h(x_n,t)$ for all $n$, their time-derivatives, and any additional entries needed to describe active processes (such as $p_\mathrm{hc}(x_n,t)$ for the one-exponential kernel). Eq. \ref{fin_jac} in Appendix~\ref{SI:jacobian} is the final result.
Diagonalizing the Jacobian $\hat{J}$ \cite{elliott2007state} yields modes as eigenvectors with corresponding eigenvalues $\lambda_j$.

\begin{figure*}
    \centering
\includegraphics[width=\linewidth]{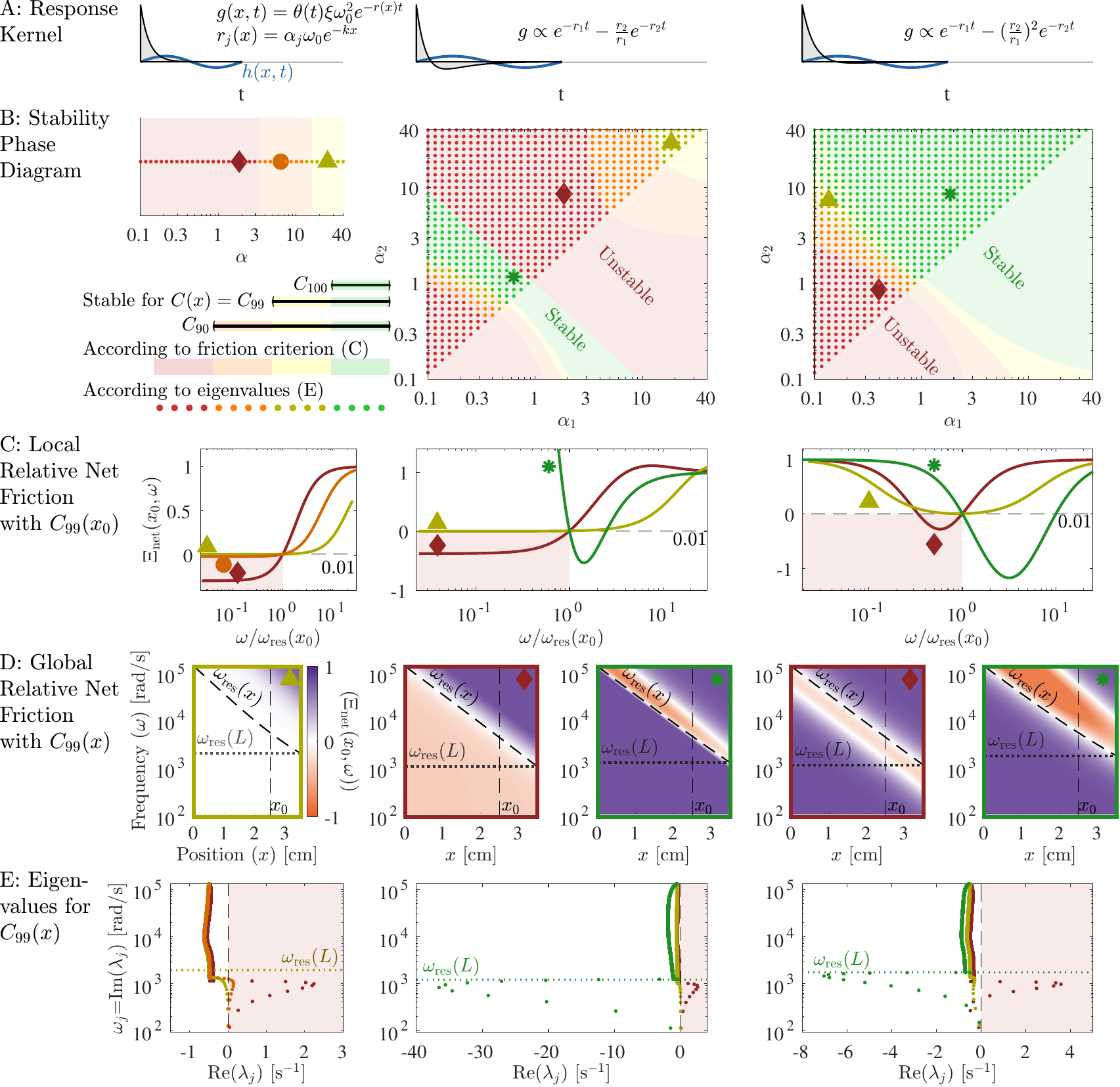}
\caption{\textbf{Extended mode stability constrains hair cell response kernels.} We consider 3 families of response kernels $g(x,\Delta t)$, with the left column showing
one exponential with a decay rate $r(x)$ adjusted by $\alpha$, Eq \ref{eq:g-exp}.
The middle and right columns have sums of two exponentials whose rates are controlled by $\alpha_{1},\alpha_{2}$, with two different choices of additive constant, Eqs~\ref{eq:perf_adap} and \ref{eq:IGkern}. 
(B) Stability phase diagrams, on which green and yellow points indicate parameters $\alpha_{1},\alpha_{2}$ at which there are no eigenvalues with positive real part when hair cells are tuned to cancel 99\% of passive friction on resonance, $C(x)=C_{99}(x)$ from Eq.~\ref{eq:C99}. If the hair cells are instead tuned to cancel 90\% of passive friction ($C_{90}$), then more cases become stable, indicated by orange points. If they cancel 100\% of passive friction ($C_{100}$), then only the green points are stable. Background colour indicates stability according to the criterion of having $\Im Z(x,\omega)>0$ for all $\omega<\omega_\mathrm{res}(x)$. Points marked by large symbols are plotted in lower panels, always with $C_{99}(x)$. (C) At a particular position $x_{0}=2.5\mathrm{cm}$, we plot the relative net friction $\relnetfriction (x,\omega)$, for frequencies $\omega$ above and below the resonance frequency $\omega_\mathrm{res}(x_{0})$. By definition of $C_{99}$, this is equal to $1-0.99$ at $\omega=\omega_\mathrm{res}(x_{0})$. The friction criterion states that any negative values at $\omega<\omega_\mathrm{res}(x_{0})$ will produce instability (red background). (D) Relative net friction for all $x$ and $\omega$. By definition of $C_{99}$, this is 0.01 along the resonance line $\omega=\omega_\mathrm{res}(x)$. Positive values (purple) indicate energy loss, while negative values (orange) indicate energy gain, potentially leading to instability. (E) Mode eigenvalues, for the same selected parameters $\alpha_{1},\alpha_{2}$ indicated by large symbols on the phase diagrams. Any eigenvalue with $\Re(\lambda_j)>0$ is an unstable mode (red background). Eigenvalues in panels B, E were calculated at $N=1000$.}
    \label{fig:big}
\end{figure*}

We find, perhaps surprisingly, that the eigenmodes fall into two qualitatively distinct classes, which we term \textit{localized} and \textit{extended} modes, whose eigenvalues and eigenvectors are plotted in Fig. \ref{fig:stab}.  
Each localized mode is strongly peaked at a specific location within the cochlea (orange and yellow in Fig.~\ref{fig:stab}B), and the location $x_\mathrm{res} < L$  of this peak is determined by its frequency $\omega_j = \Im[\lambda_j]$.
But there are a few additional eigenvalues with $\omega_j < \omega_\mathrm{res}(L)$ (blue points in  Fig.~\ref{fig:stab}A), which correspond to spatially extended eigenmodes (pink and green in Fig. \ref{fig:stab}B).
Increasing the discretization scale $N$ increases the number of localized modes, without much effect on either the 
number of extended modes present (12 in the figure), or their frequencies.

The localized modes have been studied in detail \cite{reichenbach2014physics,hudspeth2010critique,talmadge1998modeling}, and their sharpness in frequency and space is responsible for the remarkable precision with which we can sense pitch.
Numerically, we find that the $j^{th}$ mode is peaked near the resonant position of its eigenvalue, $x_\mathrm{res} \approx \log(\omega_0 / \omega_j)/k$
where $\omega_j = \lvert\Im \lambda_j\rvert$.
In the widely used and qualitatively accurate WKB approximation, the sharp peak at a given frequency is $h(x,t) \sim \lvert Z(x,\omega_j) \rvert^{-3/4}$.
Waves approaching from the oval window ($x=0$) have a decreasing wave speed as they travel right.
They slow to zero at $x_\mathrm{res}$, and deposit most of their energy in a so-called critical layer, leaving only an evanescent wave to the right of resonance~\cite{lighthill1981energy}.
Therefore, the stability of these modes is essentially determined by the stability of the local oscillator at resonance.
So long as $\Im[Z(x_\mathrm{res},\omega_j])]>0$, active processes are adding in less energy than friction is removing, and these modes are stable, $\Re(\lambda_j)<0$.
In the limit of $\Im[Z(x_\mathrm{res},\omega_j)]\rightarrow0$, these modes become infinitely peaked, and they can be thought of as essentially uncoupled oscillators acting independently, whose eigenvalues are the roots of $Z(x_n,i\lambda)=0$  (grey points in Fig. \ref{fig:stab}A).
We anticipate that these modes can thus be tuned independently and brought to the edge of instability by choosing $C(x) \approx C_{100}(x)$.

By contrast, the extended modes we find are inherently collective.
They are defined by having frequencies below the lowest resonant frequency of the cochlea, $\lvert\Im(\lambda_j)\rvert<\omega_\mathrm{res}(L)$. 
These waves travel down the entire cochlea with no evanescent region and can be thought of as sums of right- and left-moving waves that reflect off of the boundary conditions at $x=0,L$.
As with a more traditional wave equation, there is a single standing wave mode with no zero crossings (pink in Fig.~\ref{fig:stab}B), one with a single zero crossing (green in Fig.~\ref{fig:stab}B) and so on, with each successive mode having one more zero crossing and a higher characteristic frequency.
However, this pattern is cut off at a small number of crossings, corresponding to the maximum frequency still below the resonance of the helicotrema, $\omega_\mathrm{res}(L)$.  The number of extended modes, twelve for the parameters used here, thus corresponds to the number of zero crossings of the lowest-frequency localized mode. This number is set by the boundaries of the wave equation, and is independent of the discretization scale except when $N$ is very small.  While the existence of these extended modes is not a product of active processes, the active processes do influence their eigenvalues and stability. 
Interestingly, these modes are not unique features of our model and are indeed also present, albeit not discussed, in previous models \cite{neely1986model,elliott2007state,elliott2011erratum}, see Appendices~\ref{talm} and~\ref{SI:Compare}.

\subsection*{Stability of extended modes}
\noindent While the stability of localized modes is determined by the local balance of active processes and friction at resonance, the stability of extended modes is determined by a combination of these forces along the BM and the dissipation of wave energy out through the oval window.
Figure \ref{fig:stab}C shows  the relative net friction
\begin{align}
    \relnetfriction (x,\omega) = \frac{\Im[Z(x,\omega)]}{\xi\omega}
    \label{eq:neteffectivefriction}
\end{align}
as a function of frequency and position.
We observe that for frequencies below the resonant frequency of the helicotrema, $\omega< \omega_\mathrm{res} (L)$, it is negative everywhere.
As a result, energy is added to the extended modes over the entire BM, leading to their instability. This led us to ask: Are there response kernels for hair cell activity that lead to stable extended modes?
And how does requiring the extended modes to be stable limit the response kernels that hair cells might employ?

We propose a simple analytic condition that predicts whether the response kernel will destabilize the extended modes.
Fixing some $x=x_{0}$, we ask whether there is any frequency $0 \leq \omega<\omega_\mathrm{res}(x_0)$ at which $\Im[Z(x_{0},\omega)]<0$.
In that case, extended modes will experience some negative friction and may be unstable. Even though this simple criterion (shading in Fig.~\ref{fig:big}B) can be calculated without knowing the height eigenvectors or the coupling to the oval window, it predicts the stability of the extended modes (dots in Fig.~\ref{fig:big}B) well.  (See Appendix~\ref{SI:OW} for a discussion of the oval window's small effects.)
In Fig. \ref{fig:big} we use this criterion, together with the stability obtained from calculating the eigenvalues via the Jacobian, to compare three different response kernels.

\textbf{Single exponential kernel ---}
So far we have used the response kernel introduced below Eq.~\ref{eq:p_hc}, which integrates height for a time we assumed to be similar to the period of a wave resonant at that location: 
\begin{equation}
    \begin{aligned}
g(x,\Delta t) & =\theta(\Delta t)\:\xi\omega_{0}^{2}e^{-r(x)\Delta t},\quad r(x) = \alpha \passiveres(x) \\
\tilde{g}(x,\omega) & =\xi\omega_{0}^{2}\frac{r(x)-i\omega}{r(x)^{2}+\omega^{2}}.
\label{eq:g-exp}
\end{aligned}
\end{equation}
Here $\theta(\Delta t)$ is the Heaviside function that ensures a causal response.
As we have seen in Fig.~\ref{fig:stab}, for $C(x)=C_{99}(x)$ and $\alpha=2$ this choice leads to unstable extended modes since the net friction is negative everywhere for low frequencies.
At the same time, the net friction is, by construction, slightly positive at the resonant frequency.

In Fig.~\ref{fig:big}C left we plot the relative net friction at a particular location $x_0$ for several values of $\alpha$.  This $\relnetfriction(x_0,\omega)$ is monotonically increasing as a function of frequency and is given by $1-f$ at resonance, by definition.  Thus, as $f$ approaches 1, net friction always becomes negative at lower frequencies, and hence our criterion predicts instability.  In general, this criterion predicts stability only for $\alpha\gtrsim \sqrt{f/(1-f)}$, in approximate agreement with our eigenvalue results (dots in Fig.~\ref{fig:big}B).  Consequently, exponential kernels cannot bring localized modes to the edge of instability ($f\approx1$) without destabilizing extended modes, and thus we do not expect to find active processes with this form in hair cells.  Can other response kernels stabilize these extended modes?

\textbf{Approximate derivative kernel ---}
 The most common approach when studying cochlear dynamics is to assume friction is uniformly cancelled \cite{Julicher2003,hudspeth2010critique,talmadge1998modeling}, equivalent to taking a response kernel that implements an instantaneous derivative $p_\mathrm{hc}(x,t)\propto \partial_t h(x,t)$ or $\tilde{g}(x,\omega)\propto i\omega$. So if $f\leq1$, all modes at every position and frequency would experience a positive or zero relative net friction leading to stable extended modes.
However, an instantaneous derivative requires an infinitely fast response of hair cells.
With a finite response time, one can approximate a derivative by a sum of two exponentials:
\begin{equation}
g(x,\Delta t) =\theta(\Delta t)\:\xi\omega_{0}^{2}\Big[e^{-r_1(x)\Delta t}-\frac{r_2(x)}{r_1(x)}e^{-r_2(x)\Delta t}\Big]
\label{eq:perf_adap}
\end{equation}
with $r_j(x) = \alpha_j \passiveres(x)$. 
The relative coefficient ${-}r_{2}/r_{1}$ between the two terms ensures that $\int d\Delta t\:g(x,\Delta t)=0$, so that a time-independent shift in $h$ has no effect on the response.
In the limit $\alpha_1,\alpha_2\rightarrow\infty$, this response kernel approaches an instantaneous derivative.

Figure \ref{fig:big}B middle shows the stability phase diagram for this response kernel as a function of $\alpha_1$ and $\alpha_2$.
For $f<1$, the extended modes are stabilized for large $\alpha$ as in the case of a single exponential.  But there is also a narrow range with $\alpha_1 \alpha_2 \lesssim 1$ (green in Fig.~\ref{fig:big}B) where the extended modes can be stable all the way to $f=1$.
This true stability occurs because relative net friction has a negative slope at the resonance (green in Fig.~\ref{fig:big}C), so that negative net friction occurs at frequencies higher than resonance (orange in Fig.~\ref{fig:big}C).
Perhaps surprisingly, this region of true stability occurs for low rates, $\alpha_1 \alpha_2 \lesssim 1$.  Hence, hair cells could implement kernels with this form, but not in the regime where they closely approximate an instantaneous derivative.

\textbf{Zero-derivative kernel ---}
Based on our simple condition for stability, can we construct a response kernel that leads to stable extended modes for a broader range of parameters?
Since the simple analytic condition focuses on the behaviour of $\relnetfriction(x,\omega )$ at low frequencies, we approximate it by a Taylor series in $\omega$:
\begin{equation}
\label{low_om_fric}
    \relnetfriction(x,\omega )= 1+\frac{C(x)}{\xi}\Im[\left. \partial_\omega \tilde{g} \right|_{\omega=0}] + \mathcal{O}(\omega).
\end{equation}
A simple condition for positive net friction for small $\omega$ independent of $C(x)$ is thus given by $\left. \partial_\omega \tilde{g} \right|_{\omega=0}=0$, which in the time domain reads $\int_0^\infty d\Delta t \ g(x,\Delta t) \Delta t=0 $.
This condition can be met by a sum of two exponentials weighted as follows:
\begin{equation}
g(x,\Delta t) =\theta(\Delta t)\:\xi\omega_{0}^{2}\Big[e^{-r_1(x)\Delta t}-\frac{r_2(x)^2}{r_1(x)^2}e^{-r_2(x)\Delta t}\Big],
\label{eq:IGkern}
\end{equation}
where $r_j(x) = \alpha_j \passiveres(x)$ as before.
With this kernel, Fig.~\ref{fig:big} right, there are three possible cases:
In the first case, friction is positive everywhere.
This case is only possible for $f<1$ and gives rise to the yellow and orange area in the stability phase diagram.
In the second case, friction is negative in a band of frequencies below the resonance frequency, leading to instability of the extended modes, see red area in Fig.~\ref{fig:big}B right.
Finally, friction is positive below the resonant frequency but negative for a band of frequencies above the resonance frequency (green area).
Importantly, this stable regime occurs for points $\alpha_1 \alpha_2 \gtrsim 1$ and therefore does not require much tuning of parameters.

Taken together, our results suggest that the extended modes are unstable if the net friction is negative for (a band of) frequencies lower than the resonance frequency.
In principle, there are two ways to avoid this situation: either net friction is positive everywhere, or it is negative for a band of frequencies larger than the resonance frequency.
The former can only occur if the net friction at resonance isn't fully cancelled ($f<1$), thus making this regime less useful for tuning the localized modes to the edge of instability.
The latter regime also works for fully cancelled friction $f=1$.
It only occurs in a narrow parameter regime if the kernel approximates a derivative (Fig.~\ref{fig:big} middle) but can be greatly enhanced if the kernel $\tilde{g}$ instead exhibits a zero first derivative at $\omega =0$ (Fig.~\ref{fig:big} right). 
It is worth noting that for all forms of the kernel $\tilde{g}(x,\omega)$ considered here, localized modes are at least marginally stable for all $f \leq 1$.

\subsection*{Independent tuning of localized modes}
\noindent Having established conditions on the stability of extended modes, we now turn to how hair cells can tune all localized modes to the edge of instability.
\begin{figure} 
\centering
\includegraphics[width=\linewidth]{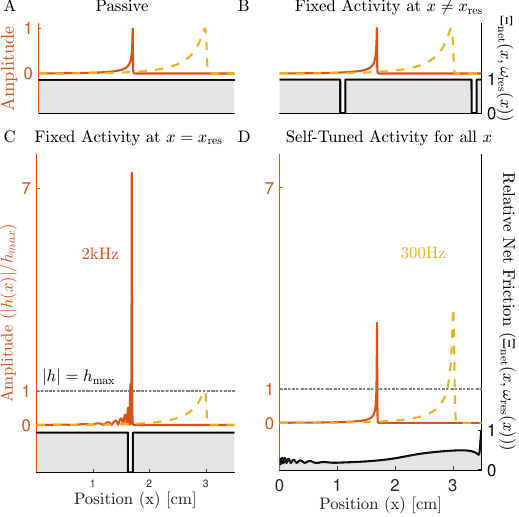}
\caption{\textbf{Canceling friction has a predominantly local effect.} We show the amplitude of the cochlea's response to pure tones at $2000$~Hz (orange) and $300$~Hz (yellow) for (A) a passive cochlea, $C(x)=0$, which by definition has $\relnetfriction(x,\omega)=1$, (B) friction reduced off-resonance, (C) friction reduced at the resonant position for $2000$~Hz, giving an over 7-fold amplification, and  (D) a cochlea self-tuned using Eq.~\ref{eq:feedback}, with target RMS height ten times greater than the passive one at each point, $h_0(x) = 10 \left.\sqrt{\left\langle h(x,t)^2\right\rangle_{\eta}} \right|_{C=0}$.
The left axis shows the amplitude of the response, normalized to peak at 1 in the passive case. The right axis shows the net effective friction at resonance $\relnetfriction(x,\omega_\mathrm{res}(x))$, Eq.~\ref{eq:neteffectivefriction}, shown in gray in all panels.
}
    \label{fig:C_sens}
\end{figure}
Since the friction term in $Z_\text{pas}$ is small compared to the canceling real parts, the relatively small absolute contribution of $Z_\text{hc}$ will dramatically affect mechanics only where the real part of $Z$ is near 0. As a result, for kernels with stable extended modes,  we propose that hair cells only need to cancel friction for the localized mode peaking at their location. Fig. \ref{fig:C_sens} shows that if we reduce friction to near zero at a specific location, we only see a noticeable amplification of frequency modes that peak near that position (Fig. \ref{fig:C_sens}C).
Off-resonance amplification  (Fig. \ref{fig:C_sens}B) looks qualitatively identical to a passive system (Fig. \ref{fig:C_sens}A). 
Tuning the cochlea therefore only requires that friction is cancelled very near the resonant frequency ($\Im[Z(x,\omega_\mathrm{res}(x))]=0$ for all $x$), a far less stringent limitation than globally nullifying friction ($\Im[Z(x,\omega)]=0$ for all $x,\omega$). 

\subsection*{Self-tuning of active process strength}
\begin{figure}
    \centering
    \includegraphics[width=\linewidth]{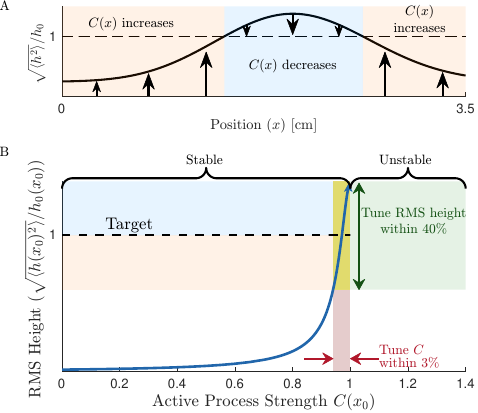}
   \caption{\textbf{Self-tuning of active processes via feedback from the order parameter (RMS height) onto the control parameter (active process strength).}  (A) A sketch of how hair cell activity is tuned to counteract friction. At positions where hair cells experience a root-mean-square height below the threshold $h_0$ (dashed line), they slowly increase their activity, reducing the effective friction via Eq. \ref{eq:feedback}, and vice versa. 
    (B) Robust tuning of hair cell activity to the critical point. At a fixed position $x_0$, we plot RMS height as a function of the activity strength $C(x_0)$ at that position. For simplicity, here we use the instantaneous derivative kernel $\tilde{g}(x,\omega)=-i\xi\omega$ for which friction is fully cancelled when $C(x)=1$. We show how a 3\% change in the activity strength $C(x_0)$ corresponds to a much larger relative change in the RMS height. Thus, while controlling $C(x_0)$ directly would require fine-tuning (red-shaded area), feedback based on the height only requires $h_0$ to fall on the steep part of this curve (green-shaded area). The region to the right of $C(x_0)=1$ is unstable.}
    \label{fig:SOC}
\end{figure}

\noindent 
So far, we have shown that requiring stability of extended modes puts constraints on viable active hair cell responses and that for stable kernel choices all localized modes can be brought to the edge of instability simultaneously.
In part this tuning is possible because the effects of $C(x)$ are local both in space and in frequency.
However, for strong amplification, it also requires the local hair cell activity strengths $C(x)$ to be tuned very close to $C_{100} (x)$, Fig.~\ref{fig:SOC}B. 
So, how can hair cells find the region where the net friction is cancelled (almost) perfectly along the line of resonant frequencies and thereby bring the set of localized modes to the proximity of their individual instabilities?
Inspired by how isolated hair cells in non-mammalian vertebrates as, for example, bullfrogs can tune themselves to their Hopf bifurcation~\cite{camalet2000auditory},
cells could take advantage of the extreme sensitivity of the membrane displacement amplitude near $C_{100}(x)$ to indirectly tune to this region, by measuring the amplitude of local displacements.
This indirect tuning works much more robustly than tuning $C(x)$ directly because the size of local oscillations in BM height are an order parameter for a dynamical bifurcation where $C(x)$ is a control parameter.

\begin{figure}
    \centering
    \includegraphics[width=\linewidth]{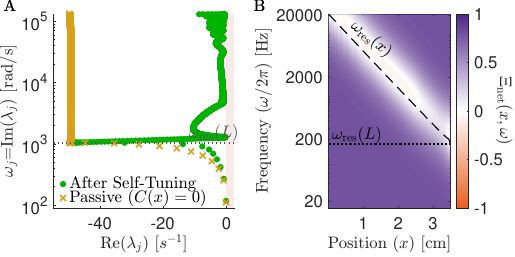}
    \caption{\textbf{Effect of self-tuning via Eq. \ref{eq:feedback}}. 
    (A) Eigenvalue structure of the passive (yellow) and the self-tuned system (green).
    Self-tuning moves all eigenvalues closer to instability.
    The localized modes near the end of the cochlea are less amplified, probably due to an interplay with the extended modes, which have most of their weight near the end of the cochlea, and boundary effects from the helicotrema.  
    Note also the presence of extended modes even in the passive system.
    (B) Relative net friction $\relnetfriction(x,\omega) = \Im Z(x,\omega)/\xi\omega$ after self-tuning. We note the slightly negative friction after resonance, similar to what is observed in the stable (green) case in Fig.~\ref{fig:big}D right.
    These plots use the zero-derivative kernel, Eq.~\ref{eq:IGkern}, with $\alpha_1=1, \alpha_2=2$.
    The target $h_0(x)$ has been set to be ten times the passive RMS height as in Fig.~\ref{fig:C_sens}D. 
    }
    \label{fig:tune}
\end{figure}

We therefore consider feedback which, rather than directly implementing $C(x) \approx C_{100}(x)$, instead adjusts $C(x)$ to target an order parameter, the RMS displacement: $\sqrt{\expval{h(x,t)^2}} \approx h_0(x)$. Because the effect of $C(x)$ is predominantly local in $x$ (Fig.~\ref{fig:C_sens}), this can be implemented by local feedback (Fig.~\ref{fig:SOC}A). Thus we consider adding additional slow dynamics to the model:
\begin{equation}
    \label{eq:feedback}
    \tau_a\frac{d C(x)}{dt}=1-\frac{\expval{h(x,t)^2}_{\eta(t)}}{h_0(x)^2}.
\end{equation}
Here we assume that the timescale of the feedback $\tau_a$ is much longer than the longest timescale of the sound-driven dynamics: $\tau_a\gg$1/20Hz.
And we imagine that the system is externally driven by input from the ear canal, and model incoming sound as uncorrelated Gaussian noise in pressure: $P_{\mathrm{ec}}(t)=\eta(t)$.
Details of the noise spectrum do not qualitatively change our results as long as there is support for all frequencies resonant on the BM (see Appendix \ref{SI:colour}). 
The temporal average needed for RMS displacement has thus been replaced by an average over this noise process --- see Appendix~\ref{SI:tuning} for details.
(Due to symmetry, the average of $h(x,t)$  vanishes, 
$\left\langle h(x,t)\right\rangle_{\eta(t)} =0$, 
thus feedback from the height squared is the first nonzero moment.)
Fig.~\ref{fig:tune} shows the eigenvalue structure and the relative net friction $\relnetfriction(x,\omega)$ of a cochlea tuned by this mechanism for a target RMS height $h_0(x)$ set to be ten times the passive RMS height. We observe that, even in this case of a relatively small target RMS height $h_0$, the real part of the eigenvalues is strongly reduced in magnitude, moving the localized modes considerably closer to the edge of instability.

More generally, this scheme can bring each mode to the edge of instability without fine-tuning any fixed parameters (green in Fig.~\ref{fig:SOC}) since $C(x)$ is a dynamical function approaching a steady-state value that only weakly depends on $h_0$.
This robust tuning is reminiscent of systems that exhibit self-organized criticality \cite{sornette_critical_1992, sornette_mapping_1995,camalet2000auditory,buendia2020feedback}, where a slowly varying control parameter is tuned via feedback from a fast order parameter.

\subsection*{Robustness to perturbations}
\begin{figure}
    \centering
    \includegraphics[width=\linewidth]{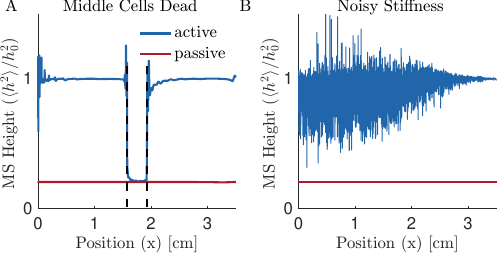}
    \caption{\textbf{The cochlear self-tuning is robust to perturbations.} Mean-square height for (A) a simulation in which the middle $10\%$ of cells are inactive, i.e. $C(x)=0$ for $0.45L<x<0.55L$, (B) a simulation for which the exponentially decaying stiffness $\omega_0^2e^{-2k x}$ has multiplicative white noise with a standard deviation of 1\%. For both plots the target $h_0(x)$ is five times the passive RMS height. Note that we use a discretization scale of $N=4000$, which might not be large enough to resolve the peaks of the high-frequency localized modes close to $x=0$ and lead to the observed spikes there in (A). }
    \label{fig:rob}
\end{figure}

\noindent  This section examines two possible sources of variability and how, despite them, the system can still find its critical point. 
First, we consider a case in which the middle 10\% of the cochlea is forced to have zero activity.
We observe that points away from this region still reach their critical points and that inactive points very close to the edges of inactivity can be partially amplified (Fig. \ref{fig:rob}A).
This observation reinforces the local nature of these active processes, showing that only friction at a given position needs to be reduced to achieve amplification for that position's resonant frequency.

Another important part of our model is the stiffness of the basilar membrane, whose position dependence determines the resonance position of the different frequencies.
To check how robust self-tuning is to slight changes in the stiffness profile, we consider a system in which the stiffness of the BM is noisy, replacing $\omega_0^2 e^{-2k x}$ with $\omega_0^2 e^{-2k x_n} (1+0.01\gamma(x_n))$, where $\gamma(x_n)$ is iid standard Gaussian noise.
Figure \ref{fig:rob}B shows the resulting steady-state mean-square height profile.
We see that self-tuning still achieves similar average enhancement, demonstrating that the exact stiffness profile is not necessary for amplification in the cochlea.
However, the resulting system is quite noisy.
We believe that this noisiness is due to the fact that the stiffness no longer decreases monotonically everywhere.
If, as a result, the real part of the impedance is negative already before resonance, the travelling wave is exponentially suppressed and decays rapidly there.
In order to match the target RMS height, hair cells at this location try to overcompensate by considerably increasing their activity.
The resulting increase in $C(x)$, however, also affects neighboring points and leads to the large spikes seen in Fig.~\ref{fig:rob}B.

\section*{Discussion}
\noindent 
The high sensitivity, dynamic range, and frequency resolution of human hearing are all thought to arise due to proximity of individual oscillators to Hopf bifurcations, driven by hair cell activity that effectively reduces friction~\cite{hudspeth2010critique}.  Here we introduce a wave equation for the basilar membrane that includes hair cell activity in terms of a generic response kernel, with a position-dependent activity strength, and analyze its mode structure. 

We find modes that peak at a resonant position, which we call localized modes, and argue that it is these modes whose Hopf bifurcations enable the fidelity of hearing.
Although different spatial locations are, in principle, coupled by fluid flow and influenced by hair cell activity throughout the cochlea, we demonstrate that for small friction, the localized modes become so sharply peaked that different positions are nearly uncoupled. 
In this small friction limit the amplification of each mode is dependent only on the activity of hair cells near the resonant position.
Thus, a simple, local feedback scheme for hair cell activity strength can tune all localized modes to the edge of instability.

Surprisingly, however, we also find a second set of modes, which we call extended modes.
These are standing-wave-like, and couple to essentially all hair cells.
The existence of these modes does not depend on hair cell activity and they are also present in our passive model.
Indeed, we also find them in previous models~\cite{neely1986model,elliott2007state,elliott2011erratum,talmadge1993new}, see Appendices~\ref{talm} and ~\ref{SI:Compare}.
We show that requiring stability of extended modes provides strong limitations on viable hair cell responses.
In particular, our results suggest that approximations to the derivative kernel, which is often implicitly assumed to underly hair cell activity~\cite{Julicher2003,wang2019emergence}, generically lead to unstable extended modes.
Other kernels might be better suited, and we find criteria for the temporal shape of active processes which hair cells must obey.  An interesting question for further research is to understand which proposed molecular mechanisms satisfy these criteria.
For kernels which obey these criteria the localized modes become unstable at smaller activity strength than the extended modes do, and so local tuning can bring the localized modes independently to the edge of instability.
In the self-organized steady-state of our model, effective friction is only cancelled along the one-dimensional resonant line through the two-dimensional space of frequency and cochlear position. This is in contrast to existing models that analyze a nonlinear cochlea by assuming that friction has been set globally to (near) zero~\cite{Julicher2003,reichenbach2014physics,talmadge1993new}.

Models for isolated hair cells without a cochlea use a feedback scheme similar to ours, with a control parameter, usually calcium activity, tuned towards a Hopf bifurcation \cite{nadrowski2004active,camalet2000auditory}.
In both cases, tuning is based on the (local) order parameter, here the BM displacement, and works robustly due to the large susceptibility at the critical point.
Using criticality for sensing might indeed be common in biological systems.
For instance, for E. coli chemosensing, we and others have proposed that cell receptor arrays tune themselves close to criticality to detect small changes in concentration \cite{keegstra2022near, sherry2024lattice}. 
In the neural realm, we have suggested that proximity to a bifurcation of the voltage dynamics might underly the incredible temperature sensitivity of pit vipers~\cite{graf2023bifurcation} and allow fruit flies to reliably extract odor timing information for navigation~\cite{choi.etal_2024}.
Furthermore, it is thought that the schooling behaviour of fish, flocking of birds and swarming of insects are near a phase transition to optimize collective computation~\cite{bonabeau1997self,niwa1994self,romanczuk2023phase}.
Finally, it has been shown that an anti-Hebbian learning rule in neurons whose connectivity is suppressed in response to activity can lead to a self-organized dynamical critical state~\cite{levina2007dynamical,magnasco2009self, landmann_self-organized_2021}.

Whether the order-parameter based feedback scheme presented here is implemented in hair cells could be tested by measuring the cochlea's response when given a continuous signal at a limited frequency bandwidth.
We expect hair cell activity at positions with a resonant frequency close to that of the signal to decrease while having a minimal effect for positions further away.
Directly testing this prediction would require measuring the response in a live cochlea, a difficult but perhaps feasible task with emerging optical coherence tomography techniques \cite{vogl2022methods,olson2020cochlear}.  It might also be possible to indirectly test this prediction with simpler psychoacoustic methods.

While our model provides a mechanism for the cochlea to tune itself using only local information, there are hints that there is at least some nonlocal feedback.
Under contralateral stimulation (sound played in the opposite ear), the frequency of known otoacoustic emissions (OAEs) shifts.
This phenomenon is thought to be due to changes in outer hair cell activity induced by signals from neurons in the MOC bundle \cite{lewis2020efferent}.
In our model, such shifts also occur naturally if contralateral stimulation globally decreases $C(x)$: If the contribution $Z_{\text{hc}}$ of hair cells to the impedance is not purely imaginary, a change in $C(x)$ shifts the resonant frequency of that position (and thereby changes the frequency of otoacoustic emissions) by a small amount $\Delta \omega \propto \Delta C\Re[\tilde{g}]$.
Such frequency shifts have indeed been observed in lizards \cite{koppl1993spontaneous}.
Should they also be observed in human hearing when stimulation happens in the same ear as the measurement of OAEs, this could lend further credence to our local order parameter-based feedback.
   
Early evidence for the importance of a Hopf bifurcation in hearing came from characteristic nonlinearities, including that the BM wave amplitude grows as the $1/3$ power of sound amplitude and a prominent third harmonic in evoked otoacoustic emissions \cite{reichenbach2014physics,hudspeth2010critique,Julicher2003}.
Our model is explicitly linear, which we expect to be a good approximation far from the edge of instability.
But when hair cell activity is strong enough to precisely bring the impedance at resonance to zero, the small nonlinearities of the basilar membrane become the dominant restoring force.
It is thus an interesting question for future research how these nonlinearities interact with extended modes.

In addition to making a linear approximation, we discretized the cochlea to understand its mode structure. This discretization separates the BM into segments of length  $L/N$ where $N$ typically ranged from 1000 to 4000 in our numerics.  The real cochlea also contains a small spatial scale, set by the length at which lateral coupling dominates, around $20\mu$m~\cite{naidu2001longitudinal,emadi2004stiffness}, corresponding to around 2000 independent segments.  We expect that the number, shapes and eigenvalues of the extended modes will be independent of details at this small spatial scale. 
 However, the details of the short length-scale physics might influence localized modes in interesting ways.

Frequency discrimination and signal amplification in the range of $20-1000$Hz remains an area of active research \cite{reichenbach2010ratchet}.
Since the extended modes exhibit resonant frequencies below the lowest resonant frequency of the basilar membrane ($165$Hz), they could potentially contribute to the cochlea's low-frequency sensitivity.
Indeed, experimental measurements of BM dynamics have revealed a constant phase slope near the cochlear apex, indicating that low-frequency waves reach the helicotrema~\cite{cooper1995nonlinear,recio2017mechanical}.
This observation aligns with the characteristics of the extended modes we present and the exploration of these extended modes and their impact on hearing continues to be an exciting avenue for future research.

\begin{acknowledgments}
\noindent
We thank Jim Hudspeth and the members of the Machta group for helpful discussions, and Pranav Kantroo, Mason Rouches, Derek Sherry and Jose Betancourt for constructive comments on the manuscript. This work was supported by the Deutsche Forschungsgemeinschaft (DFG, German Research Foundation) Projektnummer 494077061 (IRG), and by NIH R35GM138341 (BBM) and a Simons Investigator award (BBM).
\end{acknowledgments}

\bibliographystyle{ieeetr}    
\bibliography{bib}


\onecolumngrid

\titleformat{\section}{\raggedright\bfseries\large}{Appendix \arabic{section}.}{1em}{} 
\titleformat{\section}{\setcounter{equation}{0}\raggedright\bfseries\large}{Appendix \Alph{section}.}{1em}{} 
\renewcommand{\thesection}{\Alph{section}}
\setcounter{section}{0}

\renewcommand{\theequation}{\Alph{section}\arabic{equation}}
\setcounter{equation}{0}


\bigskip\

\begin{center}
\Large
\textbf{Appendices}
\normalsize
\end{center}

\noindent
The first two appendix sections show full details of how we write the discretized model in matrix form, and how we calculate the RMS height for self-tuning. The remaining sections are robustness checks of various kinds. They investigate the effect of altering parameters and the exact form of the response kernels in our model, and confirm that previous models exhibit similar behavior.


\tableofcontents

\section{Constructing the Jacobian $\hat{J}$ for linear response kernels}
\label{SI:jacobian}
\noindent  Our passive cochlear model is based on work from Ref. \cite{talmadge1998modeling}, with a few changes in notation and any larger differences discussed in Appendix \ref{talm}. We model the cochlea with two compartments. First we consider the bulk of the cochlea where sound induces flux of water in the upper and lower compartment; we define $j(x,t)=j_\text{lower}(x,t)-j_\text{upper}(x,t)$ as the difference in volume current between the scala tympani and scala vestibuli. According to force balance, this creates a corresponding pressure difference $p(x,t)=p_\text{lower}(x,t)-p_\text{upper}(x,t)$, between the two compartments, 
\begin{equation}
\label{eq:AFB}
    \rho \partial_t j(x,t)=-A_{\text{cs}} \partial_x p(x,t)
\end{equation}
where $A_{\text{cs}}$ is the average cross-sectional area of a cochlear compartment and $\rho$ is the density of water.
The fluid flux propagates down the cochlea, creating a displacement $h(x,t)$  of the BM, which we call height. The fluid flux obeys a continuity equation
\begin{equation}
    \label{eq:Acont}
2W_{\text{bm}}\partial_th(x,t)+\partial_xj(x,t)=0,
\end{equation}
where $W_{\text{bm}}$ is the width of the BM. We can eliminate $j(x,t)$ from Eq. \ref{eq:AFB} and \ref{eq:Acont} to arrive at a modified wave equation, 
 \begin{equation}
     \label{eq:Awave_eq}
     \frac{2\rho W_{\text{bm}}}{A_{\text{cs}}}\partial_t^2h(x,t)=\partial_x^2p(x,t).
 \end{equation}
 For our numerical solution, we use a finite element approximation to this equation where we break the cochlea into $N$ points separated by a distance $\delta x=\frac{L}{N}$, and we label each point $x_n=n\delta x$ with $n=1,2,...,N$. Now Eq. \ref{eq:Awave_eq} becomes,
 \begin{equation}
     \label{eq:wave_eq_disc}
     \frac{2\rho W_{\text{bm}}}{A_{\text{cs}}}\partial_t^2h(x_n,t)=\partial_x^2p(x_n,t)\approx\frac{p(x_{n+1},t)-2p(x_n,t)+p(x_{n-1},t)}{\delta x^2}.
 \end{equation}
 We now turn our attention to the boundary conditions. At the left-hand side, we have a Neumann boundary condition where the lateral displacement of the oval window $d_\mathrm{ow}(t)$ creates a flux of fluid, which propagates down the cochlea and induces an equal but opposite flux at the round window, due to fluid incompressibility.
 Via Eq.~\ref{eq:AFB}, this flux leads to a pressure gradient:
\begin{equation}
\partial_t^2{d}_{\text{ow}}(t)\approx\frac{A_{\text{cs}}}{2\rho A_{\text{ow}}}\frac{p(x_1,t)- p(0,t)}{\delta x}
\end{equation}
where $A_\mathrm{ow}$ is the area of the oval window.
The oval window itself acts as a damped harmonic oscillator:
\begin{equation}
  \label{eq:ABC}
\partial_t^2{d}_{\mathrm{ow}}(t)+\xi_{\mathrm{ow}}\partial_t{d}_{\mathrm{ow}}(t)+\omega_{\mathrm{ow}}^2d_{\mathrm{ow}}(t)=\frac{p(0,t)+G_{\mathrm{me}}P_{\mathrm{ec}}(t)}{\sigma_{\mathrm{ow}}},
\end{equation}
where  $\xi_{\text{ow}}$ is its dampening constant, $\omega_{\text{ow}}$ the middle ear resonance, $P_{\text{ec}}(t)$ the pressure in the ear canal, $G_{\text{me}}$ the gain of the middle ear, and $\sigma_{\text{ow}}$ the (2D) areal density of the oval window. At the cochlea's apical end $x=L$, a gap in the basilar membrane (the helicotrema) suggests zero pressure difference via the Dirichlet boundary condition~\cite{talmadge1998modeling}:
\begin{equation}
     p(x_N,t)=0.
 \end{equation}
Now assuming that $P_\text{ec}(t)=0$ we can write the discretized system in the form,
\begin{equation}
    \hat{F}\Vec{p}=\partial_t^2\Vec{{h}}
\label{vecwave}
\end{equation}
 We will introduce the shorthand $p(x_n,t)=p_n$ and $ h(x_n,t)=h_n$ as we write $\Vec{p}$ as an $N$ column vector,
\begin{equation}
    \Vec{p} =\begin{pmatrix}
   {p}_0\\ {p}_1 \\ \vdots\\ {p}_{N-2}\\{p}_{N-1}
\end{pmatrix}.
\end{equation}
$\Vec{h}$ is an $N$ vector of BM displacement with $h(0,t)$ replaced by $d_\text{ow}(t)$, 
\begin{equation}
    \Vec{h} = \begin{pmatrix}
   {d_{\text{ow}}}\\ {h_1} \\ \vdots\\ {h_{N-2}}\\h_{N-1} 
\end{pmatrix},
\end{equation}
and $\hat{F}$ is a modified finite difference matrix,
\begin{equation}
    \hat{F} = \frac{A_\text{cs}}{2\rho W_\text{bm}\delta x^2 }
    \begin{pmatrix}
        \frac{-\delta x W_\text{bm}}{A_\text{ow}}& \frac{\delta x W_\text{bm}}{A_\text{ow}}&0&\hdots&0 \\
        1&-2&1&\hdots&0\\
        \vdots&&\ddots& &\vdots\ \\
        0&\hdots&1&-2&1\\
        0&\hdots&0&1&-2
    \end{pmatrix}.  
\end{equation}
Note that we have used the right-hand side boundary condition ($p_N=0$) to eliminate the last row of the matrix. 

We use a modified version of the formalism from Ref. \cite{elliott2007state} to write the dynamics as
\begin{equation}
    \partial_t\Vec{X}=\hat{J}\Vec{X}
\end{equation}
where $\hat{J}$ is the system's Jacobian and $\Vec{X}$ is the state vector concatenating $d_\text{ow}(t),\partial_t d_\text{ow}(t),h(x_n,t),\partial_t h(x_n,t)$ with $n=1,2,3 ..., N-1$ and any additional degrees of freedom needed to describe active processes.
Here, we show this procedure explicitly for $g(x,\Delta t)\propto e^{-r(x) \Delta t}$.

To begin with, we write the height-pressure relation as follows,
\begin{equation}
\label{ph1}
    p(x_n,t)=(\partial_t^2h(x_n,t)+\xi\partial_th(x_n,t)+\omega_0^2e^{-2kx}h(x_n,t))\sigma_{bm}+p_\text{hc}(x_n,t)
\end{equation}
The stiffness $\omega_0^2e^{-2k x}$ is exponentially decaying in space, with $\omega_0$ denoting the resonant frequency at the base of the cochlea. $\sigma_\mathrm{bm}$ is the mass per area of the BM, and $\xi$ the friction per unit mass. 

The active hair cell contribution to the pressure is
\begin{equation}
    p_\text{hc}(x_n,t)=\sigma_{\text{bm}}\xi\omega_0^2 C(x_n)\int_{-\infty}^{\infty} d\Delta t\:\theta(\Delta t)e^{-r(x_n)\Delta t}\:h(x,t-\Delta t).
\end{equation}
for some positive function $r(x_n)$ and $C(x_n)$ defined in the main text. Direct computation of  $\partial_t p_\text{hc}(x_n,t)$ reveals,
\begin{equation}
\label{HCODE}
    \partial_t p_\text{hc}(x_n,t)=\sigma_\text{bm}\xi\omega_0^2C(x_n)h(x_n,t)-r(x_n)p_\text{hc}(x_n,t)
\end{equation}

For this particular form of $p_\text{hc}(x,t)$ we write the vector $\vec{X}$ as column vector of length $3N-1$ as follows: 
\begin{equation}
    X=\begin{pmatrix}
        \partial_t d_\text{ow}(t)\\
         d_\text{ow}(t)\\
        \partial_t h(x_1,t) \\
         h(x_1,t) \\
        p_\text{hc}(x_1,t) \\
        \vdots\\
       \partial_t h(x_{N-1},t) \\
    h(x_{N-1},t) \\
        p_\text{hc}(x_{N-1},t) \\
    \end{pmatrix}.
\end{equation}
With this choice,
\begin{equation}
\label{Emat}
   \partial_t \vec{h}=\hat{E}\vec{X},
\end{equation}
where $\hat{E}$ is a $N\times(3N-1)$ block diagonal matrix with the following form,
\begin{equation}
    \hat{E}=\begin{pmatrix}
        E_0& 0&\hdots&0&0 \\ 0&E_1&\hdots&0&0 \\ \vdots&& \ddots&&\vdots \\ 0&0&\hdots&E_{N-2}&0 \\0&0&\hdots&0&E_{N-1}
    \end{pmatrix}
\end{equation}
\begin{equation*}
    {E_0}=\begin{pmatrix}
        1& 0
    \end{pmatrix}
\end{equation*}
\begin{equation*}
    {E_n}=\begin{pmatrix}
        1& 0&0
    \end{pmatrix}
    ~~~~~~n=1,2,...,N-1
\end{equation*}
In order to get the Jacobian of $\vec{X}$ we must first take its time derivative. We can express this time derivative as a sum of a contribution from $\vec{X}$, with the prefactor of each term in $\vec{X}$ given by a matrix $\hat{D}$, and a contribution from the pressure across the BM $\vec{p}$, with the prefactor of each term in $\vec{p}$ given by a matrix $\hat{B}$,
\begin{equation}
\label{Xdot}
\partial_t\vec{X}=\hat{D}\vec{X}+\hat{B}\vec{p}.
\end{equation}
Here $\hat{B}$ is a $(3N-1)\times N$ matrix with the following form,
\begin{equation}
    \hat{B}=\begin{pmatrix}
        B_0& 0&\hdots&0&0 \\ 0&B_1&\hdots&0&0 \\ \vdots&& \ddots&&\vdots \\ 0&0&\hdots&B_{N-2}&0 \\0&0&\hdots&0&B_{N-1}
    \end{pmatrix}
\end{equation}
\begin{equation*}
    {B_0}=\begin{pmatrix}
        \sigma_\text{ow}^{-1}\\ 0
    \end{pmatrix}
\end{equation*}
\begin{equation*}
    {B_n}=\begin{pmatrix}
        \sigma_\text{bm}^{-1}\\ 0\\0
    \end{pmatrix}
    ~~~~~~n=1,2,...,N-1
\end{equation*}  
And $\hat{D}$ is a $(3N-1)\times(3N-1)$ matrix with the following form,
\begin{equation}
    \hat{D}=\begin{pmatrix}
        D_0& 0&\hdots&0&0 \\ 0&D_1&\hdots&0&0 \\ \vdots&& \ddots&&\vdots \\ 0&0&\hdots&D_{N-2}&0 \\0&0&\hdots&0&D_{N-1}
    \end{pmatrix}
\end{equation}
\begin{equation*}
    {D_0}=\begin{pmatrix}
        -\xi_\text{ow}&-\omega_\text{ow}^2\\
        1&0
    \end{pmatrix}
\end{equation*}
\begin{equation*}
    {D_n}=\begin{pmatrix}
        -\xi&-\omega_0^2e^{-2kn\delta x}&-\sigma_\text{bm}^{-1}\\ 
        1&0&0\\
        0&\sigma_\text{bm}\xi\omega_0^2C(x_n)&-r(x_n)
    \end{pmatrix}
    ~~~~~~n=1,2,...,N-1
\end{equation*}
We can then combine Eq. \ref{vecwave} with Eq. \ref{Xdot} to write,
\begin{equation}
    \dot{X}=\hat{D}X+\hat{B}\hat{F}^{-1}\ddot{h}
\end{equation}
Then substituting equation \ref{Emat} yields
\begin{equation}
    \dot{X}=\hat{D}X+\hat{B}\hat{F}^{-1}\hat{E}\dot{X}.
\end{equation}
Finally, a simple rearrangement yields,
\begin{equation}
\label{fin_jac}
\begin{split}
    \dot{X}&=\hat{J} X\\
    \hat{J}&= ( \mathbb{1} -\hat{B}\hat{F}^{-1}\hat{E})^{-1}\hat{D}
    \end{split}
\end{equation}
where $\hat{J}$ is the Jacobian of the system, a $(3N-1)\times (3N-1)$ matrix.

\begin{figure}
    \centering  \includegraphics[width=\linewidth]{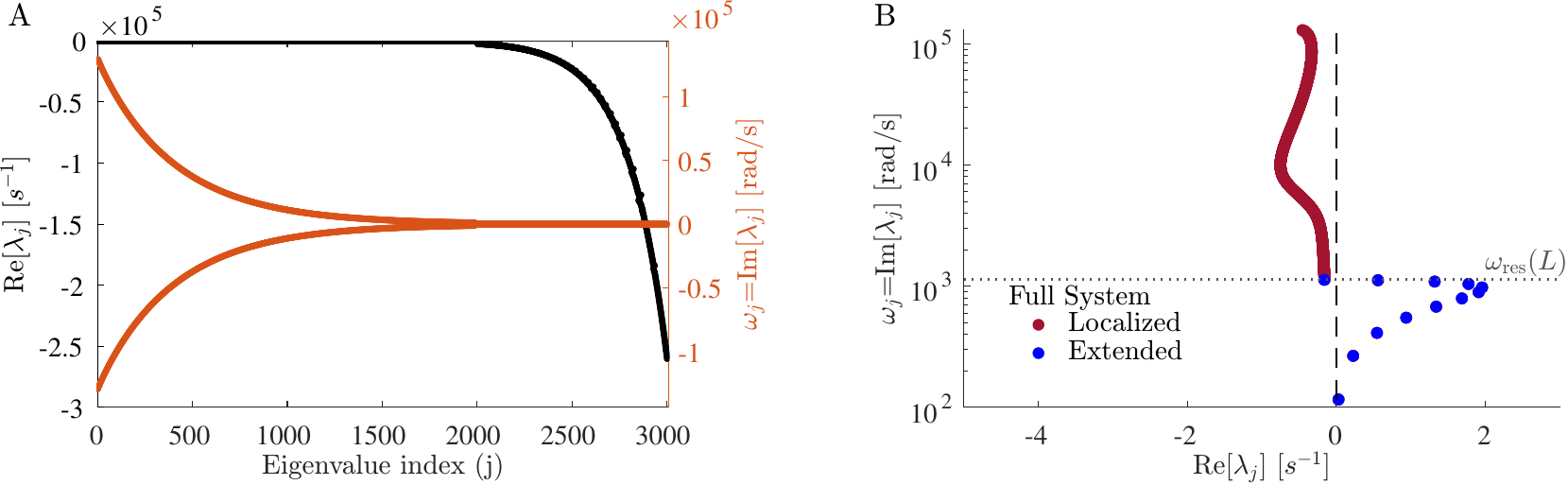}
    \caption{\textbf{All $3N-1$ eigenvalues of the Jacobian $\hat J$.}
    (A) Real (black) and imaginary (red) parts of eigenvalues $\lambda_j$, in decreasing order of $\lvert \Im \lambda_j \rvert$. Eigenvalues are in complex conjugate pairs. Those with $\Im \lambda_j = 0$ have large negative real parts, hence their modes decay quickly without oscillating.
    (B) For the same data, we plot only the eigenvalues with $\Im \lambda_j > 0$. This is the same plot as Fig.~\ref{fig:stab}A in the main text.
    For both plots $g(x,\Delta t)\propto e^{-2\omega_\text{res}^\text{pas}(x) \Delta t}$ with  99\% of friction canceled at resonance ($C(x)=C_{99}(x)$) and $N=1000$.
    }
    \label{fig:w3}
\end{figure}
\noindent

This Jacobian has $3N-1$ eigenvalues. $2N-2$ of these come from the position and velocity of the basilar membrane and are oscillating modes. The eigenvectors and corresponding eigenvalues of these modes are the localized and extended modes discussed in the main text. $2$ modes correspond to predominately oval window motion. The remaining $N-1$ modes come from active processes and are non-oscillating (imaginary part of 0). These modes have an eigenvalue with a real part far less than 0; and are absent if active processes are excluded. Fig \ref{fig:w3} shows all $3N-1$ eigenvalues, for a kernel $g(x,\Delta t)\propto e^{4\omega_\text{res}^\text{pas}(x) \Delta t}$. 
Everywhere else, we exclude eigenvalues with $\Im\lambda_j=0$ as these have a real part 4-5 orders of magnitude lower than those of oscillating modes and will quickly decay to 0.
And we exclude eigenvalues with $\Im\lambda_j<0$, as these are simply complex conjugates to the ones with $\Im\lambda_j>0$.

For other response kernels we must change the way activity is modeled. In the approximate derivative and zero derivative kernel, for example, $\hat{J}$ becomes, a $(4N-1)\times (4N-1)$ matrix and $\vec{X}$ now has a term $p_{\text{hc1}}$ and $p_{\text{hc2}}$ for each exponential contribution, respectively. The exact nature of these changes will depend on the equivalent equation to  Eq. \ref{HCODE} but in general each additional degree of freedom will increase the size of $\hat{J}$ by N.

\section{Calculating the expected height squared $\expval{h(x,t)^2}_{\eta(t)}$ for self-tuning}
\label{SI:tuning}
\noindent
In the main text, we introduce a tuning scheme in which the strength of active processes changes with the variance of BM displacement. This section explains how we calculate $\expval{h^2}$. We start by writing Eq. \ref{eq:Awave_eq} and \ref{ph1} in Fourier time,
\begin{equation}
\label{FTwave}
    -\omega^2 \frac{2\rho W_{\text{bm}}}{A_{\text{cs}}} \tilde{h}(x,\omega)=\partial_x^2\tilde{p}(x,\omega)
\end{equation}
\begin{equation}
\label{FTimped}
\tilde{p}(x,\omega)=Z(x,\omega)\tilde{h}(x,\omega)\sigma_\text{bm}=(-\omega^2+i\xi\omega+\omega_o^2e^{-2k x})\tilde{h}(x,\omega)\sigma_\text{bm}+C(x)\tilde{g}(x,\omega)\tilde{h}(x,\omega)\sigma_\text{bm}
\end{equation}
where a tilde represents a temporal Fourier transform, ($\tilde{\phi}(\omega)=\int dt e^{-i\omega t}\phi(t)$).
We want to discretize  the system into N segments, each segment obeying,
\begin{equation}
\tilde{p}_n(\omega)=Z(x_n,\omega)\tilde{h}(\omega)\sigma_\text{bm}=(-\omega^2+i\xi\omega+\omega_o^2e^{-2k x_n}+C(x_n)\tilde{g}(x_n,\omega))\tilde{h}_n \sigma_\text{bm}~~~~~~~~n=0,1,...,N-1.
\end{equation}
Using a discrete derivative,
\begin{equation}
\tilde{p}''_n=\frac{\tilde{p}_{n+1}-2\tilde{p}_n+\tilde{p}_{n-1}}{\delta x^2}, 
\end{equation}
we rewrite the right-hand side boundary conditions in Fourier time,
\begin{equation}
\label{OWeom}
\tilde{d}_{\text{ow}}(-\omega^2+i\omega \xi_{\text{ow}}+\omega^2_{\text{ow}})=\tilde{d}_{\text{ow}}Z_{\text{ow}}=\frac{1}{\sigma_{\text{ow}}}(\tilde{p}_0+G_{\text{me}}\tilde{P}_{\text{ec}}(\omega))
\end{equation}
\begin{equation}
\frac{\tilde{p}_1-\tilde{p}_{0}}{\delta x}=-\omega^2\tilde{d}_{\text{ow}}\frac{2\rho A_{\text{ow}}}{A_{\text{cs}}}.
\end{equation}
Then we use this equation to define $\tilde{p}_0$ in terms of $\tilde{d}_{\text{ow}}$ and $\tilde{p}_1$ and eliminate $\tilde{p}_0$ in Eq \ref{OWeom}. We assume $\tilde{P}_{\text{ec}}=\eta(\omega)$  where $\eta(\omega)$ is Gaussian white noise $\expval{\eta(\omega)\eta(\omega')}=\sigma^2\delta_{\omega,-\omega'}$. 

It is easiest to work with pressure, so we use Eq. \ref{FTimped} to write height in terms of pressure difference. Then, we can write this system in matrix form. 
\begin{equation}
    M_\omega \begin{pmatrix}\tilde{d}_{\text{ow}} \\ \tilde{p}_1\\
\vdots\\ \tilde{p}_{N-1} \\ \end{pmatrix}
=\begin{pmatrix} G_\text{me}\eta(w) \\ 0\\
\vdots \\ 0\\ \end{pmatrix}
\end{equation}
\small
\begin{equation*}
M_\omega=
    \begin{pmatrix}
    Z_{\text{ow}}\sigma_{\text{ow}}-\frac{2\rho\omega^2 A_{\text{ow}}\delta x}{A_\text{cs}}& -1&0&0&\hdots& 0\\
    \frac{-2\rho\omega^2{A_\text{ow}}}{A_\text{cs}\delta x}&\frac{1}{\delta_x^2}-\frac{\omega^22\rho W_\text{bm}}{A_\text{cs}\sigma_\text{bm}Z(\delta x,\omega)} &\frac{-1}{\delta x^2}& 0&\hdots& 0\\
    0&\frac{-1}{\delta_x^2} &\frac{2}{\delta_x^2}-\frac{\omega^22\rho W_\text{bm}}{A_\text{cs}\sigma_\text{bm}Z(2\delta x,\omega)}& \frac{-1}{\delta_x^2}&\hdots& 0\\
    \vdots &&&\ddots\\
     0&0&\hdots&\frac{-1}{\delta_x^2} &\frac{2}{\delta_x^2}-\frac{\omega^22\rho W_\text{bm}}{A_\text{cs}\sigma_\text{bm}Z((N-2)\delta x,\omega)}& \frac{-1}{\delta_x^2}\\
    0&0&\hdots&0&\frac{-1}{\delta_x^2} &\frac{2}{\delta_x^2}-\frac{\omega^22\rho W_\text{bm}}{A_\text{cs}\sigma_\text{bm}Z((N-1)\delta x,\omega)}\\
    \end{pmatrix}
\end{equation*}
\normalsize
To find the pressure at any given location, we have to invert $M_\omega$ (one can make use of the tridiagonal structure of $M_\omega$ for a computational speed-up \cite{tri}).
\begin{equation}
    \tilde{p}_n=[M_\omega^{-1}]_{n+1,1}G_{\text{me}}\eta(\omega)
\end{equation}
\begin{equation}
    \tilde{h}_n=\frac{[M_\omega^{-1}]_{n+1,1}}{Z(n\delta x,\omega)}\frac{G_{\text{me}}\eta(\omega)}{\sigma_{\text{bm}}}
\end{equation}
We can then make use of the Wiener-Khinchin theorem, which states,
\begin{equation}
    \expval{\Tilde{x}(\omega)\Tilde{x}(\omega')}=2\pi\delta(\omega+\omega')\tilde{\tau}(\omega)
\end{equation}
where $\tilde{\tau}(\omega)$ is the Fourier transform of the system's autocorrelation function. Applying this theorem to our systems yields for the variance of the height
\begin{equation}
    \tau(t=0)=\langle h_n^2\rangle=\frac{1}{2\pi}\int d\omega\langle h_n(\omega)h_n(-\omega)\rangle=\frac{\sigma^2G_{\text{me}}^2}{2\pi\sigma_{\text{bm}}^2}\int d\omega\abs{\frac{[M_\omega^{-1}]_{n+1,1}}{Z(n\delta x,\omega)}}^2.
\end{equation}
We then use this to iterate feedback on $C(x_n)$ of the form,
\begin{equation}
    C(x_n,t+\Delta t)=(1-\frac{\expval{h_n^2}}{h_0^2(x_n)})\Delta t +C(x_n,t).
\end{equation}
It is worth noting that this feedback assumes an infinite separation in time scale between the average $\langle h_n^2\rangle$ and the feedback on $C(x)$.

\section{Model variants in Talmadge et. al. \cite{talmadge1998modeling}}\label{talm}
\begin{figure}
    \centering
    \includegraphics[width=\linewidth]{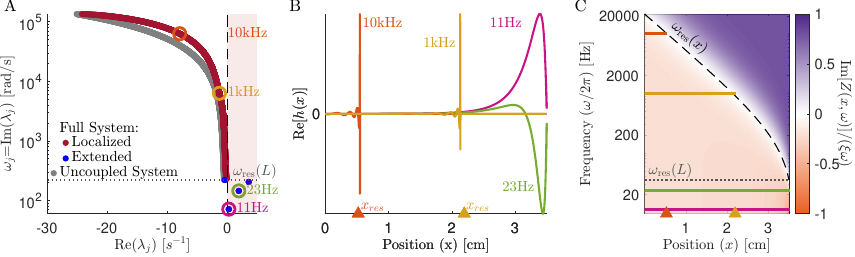}
    \caption{
        \textbf{Eigenmodes for another variant model from Talmadge et. al \cite{talmadge1998modeling}.}
        Compared to Fig.~\ref{fig:stab} in the main text, we restore their constant $\omega_{1}$
        in passive resonance $\omega_{\mathrm{res}}^{\mathrm{pas}}(x)=\omega_{0}e^{-kx}+\omega_{1}$,
        and replace our constant friction $\xi$ with their $(\gamma_{0}+\gamma_{1}e^{-k_{\gamma}x})$,
        assuming $k_{\gamma}=k$.
        The constant $\omega_{1}$ affects the low-frequency modes, reducing the number of extended modes to just four.
        The decaying friction from $\gamma_{1}$ makes the high-frequency modes more stable.
        The active model used is the same as Fig.~\ref{fig:stab}, with
        response kernel $g(x,\Delta t)\propto e^{-2\omega_{\mathrm{res}}(x)}$
        and strength $C(x)=C_{99}(x)$, which again makes the extended modes unstable. 
        Discretization uses $N=1000$ elements.
    }
    \label{fig:talm}
\end{figure}

\noindent
The model described above is taken from Talmadge et. al., Ref.~\cite{talmadge1998modeling},
with some simplifications. In fact the authors of \cite{talmadge1998modeling} study several
variants: They begin with a nonlinear model, with a particular form
of active feedback, and with non-constant friction. But they simplify
the passive model as they proceed, and replace active feedback with
an assumption that friction is perfectly canceled. This section summarizes
how these variants differ, and Fig.~\ref{fig:talm} shows that extended modes still
exist even if we pick a different variant.

They write $\xi(x,t)$ for the displacement of the BM, which in our
notation is $h(x,t)$. Their $P_{d}(x,t)$ is the pressure difference
we call $p(x,t)$. Their Eq.~(2) is very similar to our Eq.~(8), but
is written in the time domain:
\begin{equation}
P_{d}(x,t)/\sigma_{\mathrm{bm}}=\partial_{t}^{2}h(x,t)+\gamma_{\mathrm{bm}}\:\partial_{t}h(x,t)+\omega_{\mathrm{bm}}^{2}\:h(x,t).
\end{equation}
The first term is the same inertial term we have, our $-\omega^{2}\tilde{h}$
in the frequency domain. The second term includes our friction term
$i\omega\tilde{h}$ (our constant $\xi$ is their $\gamma_{0}$) but
adds a spatially varying friction $\gamma_{1}$, and a cubic nonlinearity
$\gamma_{2}$:
\begin{equation}
\gamma_{\mathrm{bm}}=\gamma_{0}+\gamma_{1}e^{-k_{\gamma}x}+\gamma_{2}(x)\:h(x,t)^{2},\qquad\gamma_{2}(x)=\frac{\gamma_{0}+\gamma_{1}e^{-k_{\gamma}x}}{b_{\mathrm{nl}}^{2}}.
\end{equation}
The third term includes our decaying stiffness $\omega_{0}e^{-kx}$
(our $k$ is their $k_{\omega}$) but adds a constant $\omega_{1}$,
and fast/slow feedback terms $\kappa_{f},\kappa_{s}$:
\begin{equation}
\omega_{\mathrm{bm}}^{2}=\big(\omega_{0}e^{-kx}+\omega_{1}\big)^{2}+\kappa_{f}(x)\:h\big(x,t-\tau_{f}(x)\big)+\kappa_{s}(x)\:h\big(x,t-\tau_{s}(x)\big).
\end{equation}
In our setup, these feedback terms are contributions to the active
pressure $p_{\mathrm{hc}}$, our Eq.(~\ref{eq:p_hc}), with delta-function kernels
at delays of $\tau_{f}(x)$ and $\tau_{s}(x)$: 
\begin{equation}
C(x)g(x,\Delta t)=\kappa_{f}(x)\:\delta\big(\Delta t-\tau_{f}(x)\big)+\kappa_{s}(x)\:\delta\big(\Delta t-\tau_{s}(x)\big).
\end{equation}
Their fast and slow time-scales are $\tau_{f}(x)=0.24\times2\pi/\omega_{\mathrm{res}}^{\mathrm{pas}}(x)$
and $\tau_{s}(x)=1.742\times2\pi/\omega_{\mathrm{res}}^{\mathrm{pas}}(x)$,
and the prefactors $\kappa_{f},\kappa_{s}$ are both positive. This
scaling of delay $\Delta t$ with the resonant frequency is the same
as for our kernels $g(x,\Delta t)$.

However, they simplify this initial model later in their paper. In
order to study the WKB approximation in their section C3, they set
friction to zero instead of keeping track of activity (i.e., $\gamma_{\mathrm{bm}}=0$)
and simultaneously set $\omega_{1}=0$, before their Eq. (58). This
brings them to the same model as we consider in the main text, with
an instantaneous derivative kernel $\tilde{g}(x,\omega)\propto i\omega$
tuned to $C_{100}(x)$.

Our model in the main text keeps friction, and studies several continuous
response kernels $g(x,\Delta t)$, while setting to zero their parameters
$\gamma_{1}$, $\gamma_{2}$, $\omega_{1}$. What happens if we restore
some of these features? In the passive model:
\begin{itemize}
\item The effect of $\omega_{1}=-145.5\times2\pi$Hz is to shift the passive
resonant frequency down to $\omega_{\mathrm{res}}^{\mathrm{pas}}(x)=\omega_{0}e^{-kx}+\omega_{1}$.
This is most important at low frequencies, at the end of the cochlea.
It reduces $\omega_{\mathrm{res}}^{\mathrm{pas}}(L)/2\pi$ from $165$Hz
to $19.5$Hz.
\item The effect of $\gamma_{1}=100\text{s}^{-1}$ is a little unclear,
as the constant $k_{\gamma}$ does not appear to be specified anywhere.
If we assume that it is equal to $k_{\omega}$ (our $k$, from the
scaling of stiffness) then this decaying term is important only at
high frequencies.
\item The effect of $\gamma_{2}$ is not something our linear eigenmode
analysis can access. 
\end{itemize}
Figure~\ref{fig:talm} shows the effect of nonzero $\omega_{1}$ and $\gamma_{1}$
on Fig.~\ref{fig:stab} of the main text. It uses the same single-exponential kernel
$g(x,\Delta t)\propto e^{-2\omega_{\mathrm{res}}(x)}$ as before,
and again cancels 99\% of passive friction at resonance, $C(x)=C_{99}(x)$.
We observe that this model still has extended modes, although not
as many of them, and at lower frequencies: 11, 21, 32, and 35 Hz.
We also observe that high-frequency modes are more stable (larger
negative $\mathrm{Re}(\lambda)$) than in Fig.~\ref{fig:stab}; this effect survives
even after tuning to $C_{99}(x)$.

\begin{figure}
    \centering
    \includegraphics[width=\linewidth]{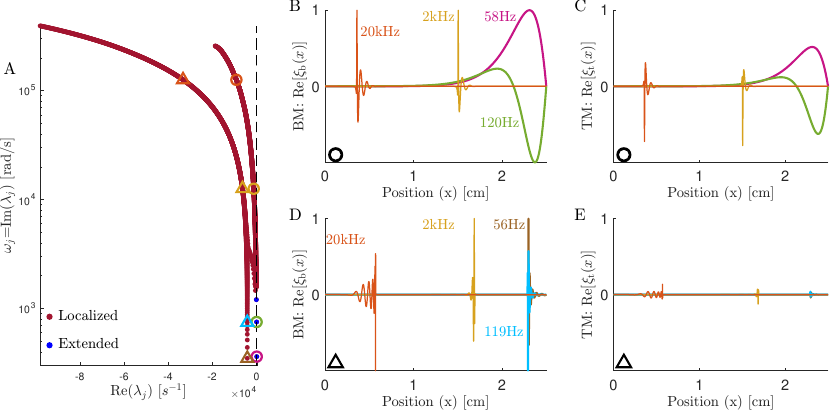}
    \caption{
        \textbf{Extended modes in the model of a cat cochlea from Refs. \cite{elliott2007state,neely1986model}.} 
        (A) Eigenvalues for activity level $\gamma=0.85$, the same as
        Fig 7(b) of Ref.~\cite{elliott2007state} except plotted with frequency on a log scale. The
        blue points are isolated eigenvalues, while the red points approach
        a continuum as $N$ is increased; the plot shows $N=500$.
        (B,C) Eigevectors
        of the circled eigenvalues, showing both the BM displacement $\xi_{b}$,
        which is $h$ in our model, and tectorial membrane (TM) displacement
        $\xi_{t}$. In addition to the familiar localized modes, we highlight
        the presence of 3 extended modes, with 58 and 120Hz plotted. (D,E)
        Eigenvectors corresponding to the points within triangles, taken from the left
        line of eigenvalues in panel A. At similar frequencies, the peak BM
        displacement in panel D is displaced in $x$ from that in panel B.
        The lowest-frequency mode of this series, 56Hz, is still a
        sharply peaked localized mode.
    }
    \label{fig:EV_elliott}
\end{figure}

\section{Comparison to the model of Neely \& Kim \cite{neely1986model}}\label{SI:Compare}
\noindent
The way we set up the Jacobian of the linearized system is similar
to what was done by Elliott, Ku \& Lineton \cite{elliott2007state}, and this section
looks at precisely their example --- a model of the cochlea of a
cat. We find that their eigensystem does have extended modes, shown
in Fig.~\ref{fig:EV_elliott}. This model was taken from Neely \& Kim \cite{neely1986model}, and we
begin by reviewing this, and comparing to our model.

The model of Neely \& Kim \cite{neely1986model} has two dynamical variables at
every position $x$ along the length of the cochlea, $x\leq2.5$cm.
These are $\xi_{b}(x,t)$, the displacement of the BM, and $\xi_{t}(x,t)$,
the displacement of the tectorial membrane (TM). They also work with
the following scaled combinations, with $b$ and $g$ fixed geometric
factors:
\begin{align}
\xi_{p}(x,t) & =b\:\xi_{b}(x,t), \qquad b=0.4\\
\xi_{c}(x,t) & =g\:\xi_{b}(x,t)-\xi_{t}(x,t), \qquad  g=1.
\end{align}
The pressure difference $P_{d}$ across the BM is coupled to scaled
BM displacement $\xi_{p}$ by their Eq. (1) (here $H$ is our $A_{\mathrm{cs}}/W_{\mathrm{bm}}$,
the height of the scala vestibuli): 
\begin{equation}
\partial_{x}^{2}P_{d}(x,t)=\frac{2\rho}{H}\partial_{t}\xi_{p}(x).
\end{equation}
Their active pressure $P_{a}$ responds to the difference $\xi_{c}$
via an impedance $Z_{4}$, which enters with the wrong sign for friction,
what they call ``negative damping''.
All in all, there are four impedances,
appearing in their Eq. (9--11). In the frequency domain, replacing
$\dot{\xi}_{b}$ with $i\omega\tilde{\xi}_{b}$ etc., these equations can
be written:
\begin{align}
\tilde{P}_{d}-\tilde{P}_{a} & =gZ_{1}i\omega\tilde{\xi}_{b}+Z_{3}i\omega\tilde{\xi}_{c}\\
0 & =Z_{2}i\omega\tilde{\xi}_{t}-Z_{3}i\omega\tilde{\xi}_{c}\\
\tilde{P}_{a} & =-\gamma Z_{4}i\omega\tilde{\xi}_{c}.
\end{align}
Each impedance $Z_{n}$ has stiffness, inertia, and friction terms
(compare our Eq. 8, noting that their $i\omega Z_{n}$ follows similar
conventions to our $Z$), with $m_{3}=m_{4}=0$:
\begin{equation}
i\omega Z_{n}(x,\omega)=k_{n}(x)-\omega^{2}m_{n}(x)+i\omega c_{n}(x),\qquad n=1,2,3,4.
\end{equation}
The factor $-\gamma$ controls the strength of the active processes.
The system becomes unstable when $\gamma>1$. The term $-\gamma i\omega c_{4}$
is active anti-friction, an instantaneous derivative kernel, but the
entire effect of $\gamma>0$ is more complicated.

Eliminating $\tilde{\xi}_{t}$ and $\tilde{P}_{a}$ from these three
equations, they can write the relation between BM displacement and
pressure difference in terms of an effective ``partition impedance''
$Z_{p}$, their Eq. (6):
\begin{equation}
\tilde{P}_{d}(x,\omega) =i\omega Z_{p}(x,\omega)\:\tilde{\xi}_{b}(x,\omega),
\qquad
Z_{p} =\frac{g}{b}\left[Z_{1}+Z_{2}\frac{Z_{3}-\gamma Z_{4}}{Z_{2}+Z_{3}}\right]
\end{equation}
Naively, it looks like $i\omega Z_{p}$ should be comparable to our
$Z_{\mathrm{pas}}+Z_{\mathrm{hc}}$, with $\gamma=0$ giving the passive
term. However, setting $\Re(i\omega Z_{p})=0$ does not predict the
correct resonant locations. 

Figure \ref{fig:EV_elliott} shows the eigenvalues of this model. Panel A is precisely Fig.~7(b) of Ref.~\cite{elliott2007state}, except that we plot the frequencies on a log scale.
We observe  some isolated points at low frequencies (blue points),
as well as the same two lines of eigenvalues seen in Ref.~\cite{elliott2007state} (red
points), which fill in a continuum as $N$ is increased. Panels B and C
plot the eigenvectors for selected eigenvalues from the right-hand line (marked with circles), and shows that
the isolated eigenvalues do indeed look like extended modes.

The resonant position of localized modes in this model is confusing.
As in Fig 7(b) of ~\cite{elliott2007state}, the eigenvalues form into two lines, hence
there are two modes near to any given driving frequency. Comparing
Fig.~\ref{fig:EV_elliott}B (modes from the right-hand line, indicated by circles) and
Fig.~\ref{fig:EV_elliott}D (left-hand line, triangles), we see that the two modes at about
20kHz excite the BM at quite different locations. None of the plots
in Ref.~\cite{elliott2007state} show eigenvectors.

We note that our plots use parameters as corrected in 2011, Ref.~\cite{elliott2011erratum}.
Using the parameters from the original paper \cite{elliott2007state} (which itself
claimed to correct typos in Neely \& Kim \cite{neely1986model}) gives an eigenvalue
spectrum with only one extended mode (not shown).

\section{Effect of discretization scale $N$ and activity strength $C_{100f}(x)$ on mode structure}
\label{SI:scaling}
\begin{figure}
    \centering
    \includegraphics[width=\linewidth]{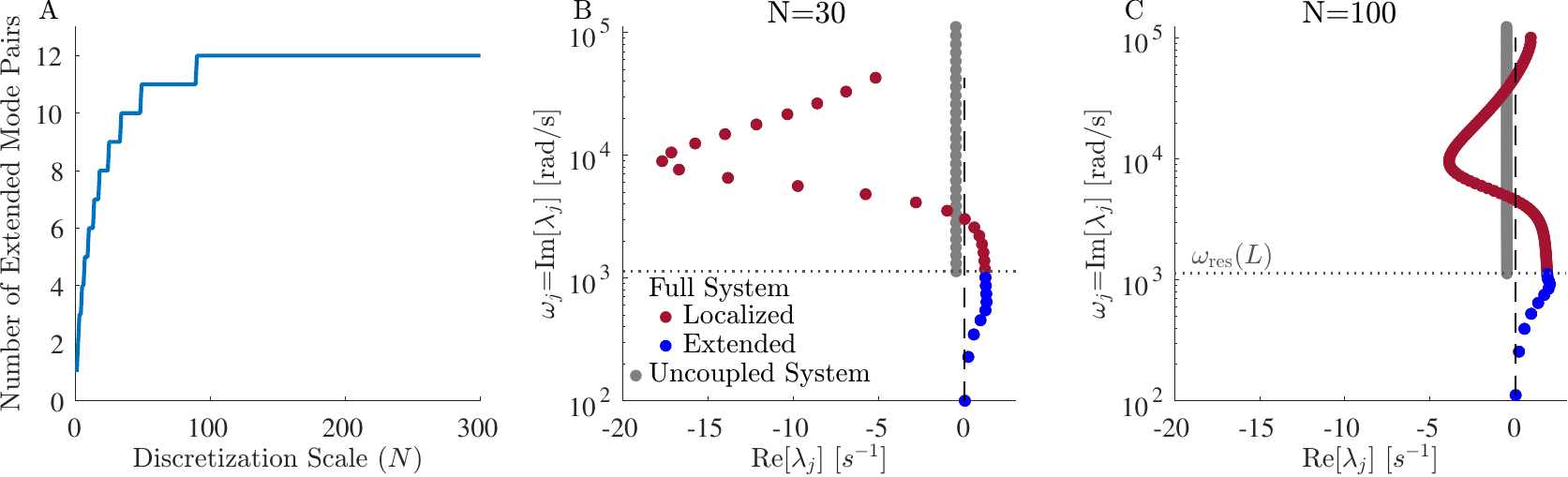}
    \caption{\textbf{Effect of discretization scale $N$ on the extended and localized modes.}
        (A) The number of extended modes as a function of $N$, showing a plateau at 12 which is reached at $N=101$. (B,C) Eigenvalue structure for $N=30$ and $N=100$,
        showing that the spacing between localized modes decreases as $N$ is increased.
        Panels B  and C are identical to Fig.~\ref{fig:stab}A except for using smaller $N$.
    }
    \label{fig:Nscale_ext}
\end{figure}

\noindent
Here, we show how both the localized modes and extended modes scale with discretization and friction. We claim that the extended modes are independent of the discretization scale; this is true unless $N$ is small. Fig.~\ref{fig:Nscale_ext}A shows how the number of extended modes on the basilar membrane scales with $N$. One can note that the number of these modes plateaus at $N=101$. For values of $N$ larger than this, 12 modes will remain for simulations run with our parameter values. This makes sense because for values of $N<101$ the discretization scale $\delta x$ is larger than the geometric wave number of the cochlea $k_o=\sqrt{\frac{2\rho W_{\text{bm}}}{A_\text{cs}\sigma_\text{bm}^2}}$ and the finite element model is a poor approximation. We can also see that increasing the discretization scale decreases the spacing between localized modes in Fig. \ref{fig:Nscale_ext}B,C. Note that some localized modes appear unstable in Fig. \ref{fig:Nscale_ext}B,C, this is a small-$N$ effect, Fig. \ref{fig:w3}B shows the same system with $N=1000$ where they are all stable.

\begin{figure}
    \centering
    \includegraphics[width=\linewidth]{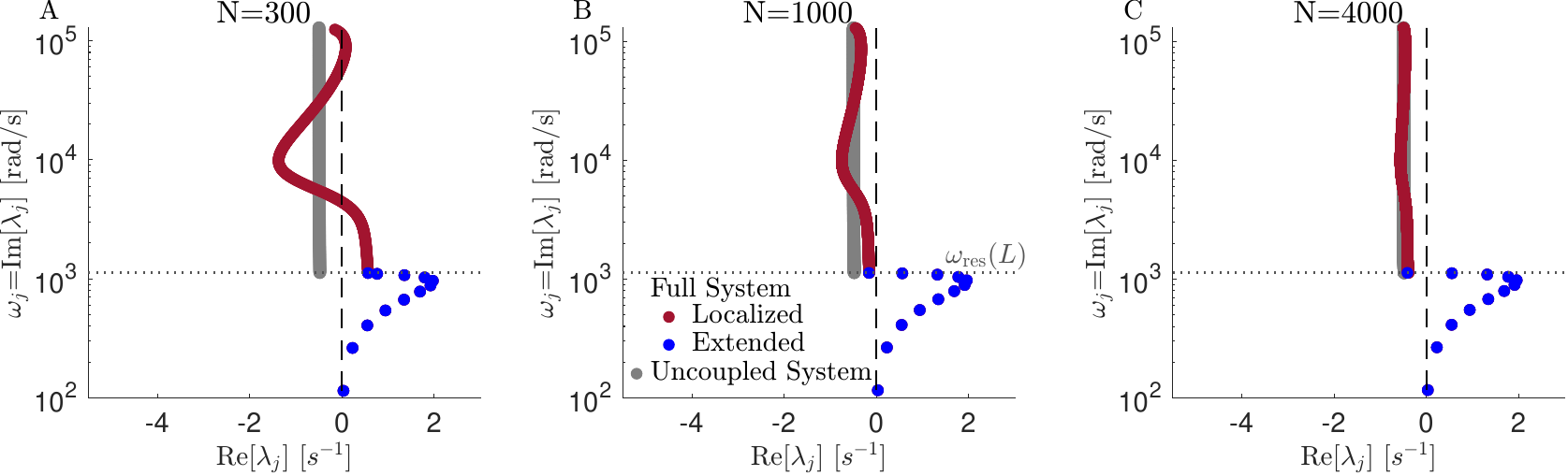}
    \caption{\textbf{Effect of discretization scale $N$ on the localized modes.}
    We observe that the localized mode eigenvalues become much closer to those of theuncoupled system as we increase the discretization scale: (A) $N=300$, (B) $N=1000$, (C) $N=4000$. 
    Note that a discretization scale of $N=4000$ has a spacing smaller than the measured persistence length of the basilar membrane.
    These plots are identical to Fig.~\ref{fig:stab}A except for varying $N$.}
    \label{fig:Nscale_loc}
\end{figure}

\begin{figure}
    \centering
    \includegraphics[width=\linewidth]{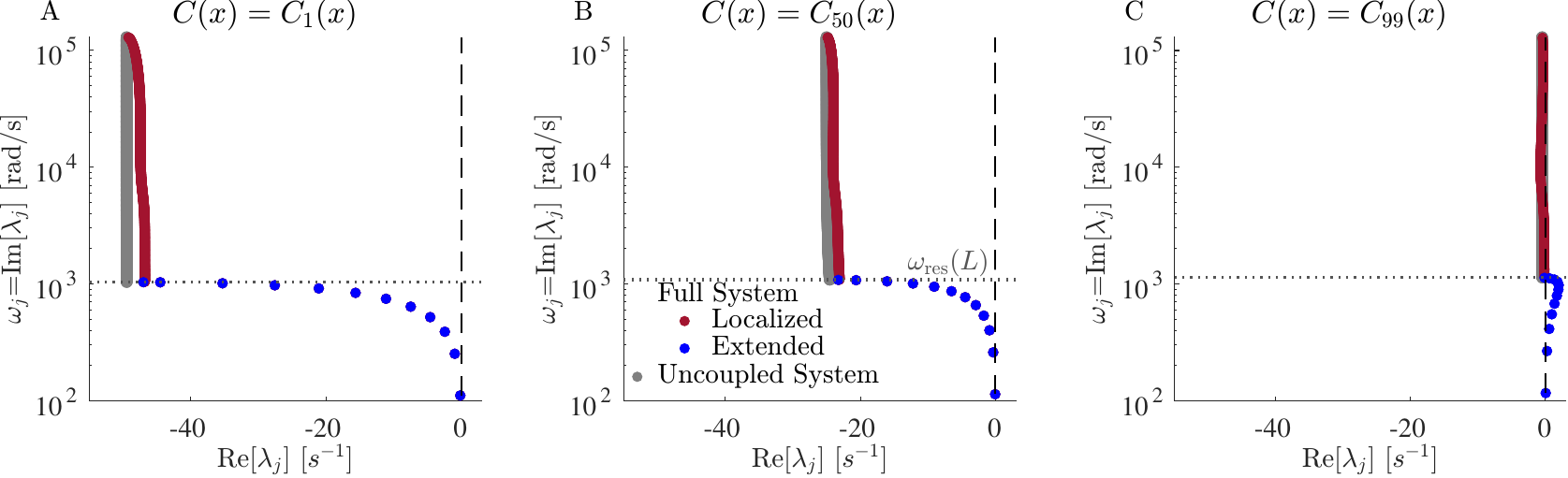}
    \caption{\textbf{Mode structure when cancelling 1\%, 50\% or 99\% of passive friction.}  We observe that the localized mode eigenvalues become closer to those of the uncoupled system as we increase the proportion of friction cancelled at resonance. (A) 1\% canceled, (B) 50\% canceled, (C) 99\% canceled. The cochlea is underdamped, so even with friction 1\% cancelled, the localized modes are similar to the uncoupled ones. These plots are otherwise like Fig.~\ref{fig:stab}A, using exponential feedback with $r(x)=2\omega_\text{res}^\text{pas}(x)$, and $N=1000$.}
    \label{fig:xiscale_loc}
\end{figure}

We can also see how the discretization changes the behaviour for larger values of N. Here, we demonstrate that for low friction and large $N$, the localized modes approach the modes of a spatially uncoupled cochlea (grey dots showing eigenmodes of the matrix $\hat{D}$), justifying the claim that in the low friction and large $N$ limit localized modes act effectively independently. Fig \ref{fig:Nscale_loc} shows that if friction is 99\% cancelled, increasing $N$ makes localized modes qualitatively indistinguishable from the uncoupled modes. Note that they cannot correspond exactly as there are $12$ less localized modes than uncoupled ones. We see a similar trend in Fig. \ref{fig:xiscale_loc} as we increase $f$. It is worth noting that even with only 1\% of friction at resonance cancelled, the localized modes are not too dissimilar from the uncoupled modes as the fully passive cochlea already has low friction. If we make $N$ larger at low $f$, we see a similar convergence, though localized and uncoupled modes are always further apart than in an identical system with larger $f$.

\section{Variations to the response kernels $g(x,\Delta t)$}
\label{SI:kernels}
\begin{figure}
    \centering
    \includegraphics[width=\linewidth]{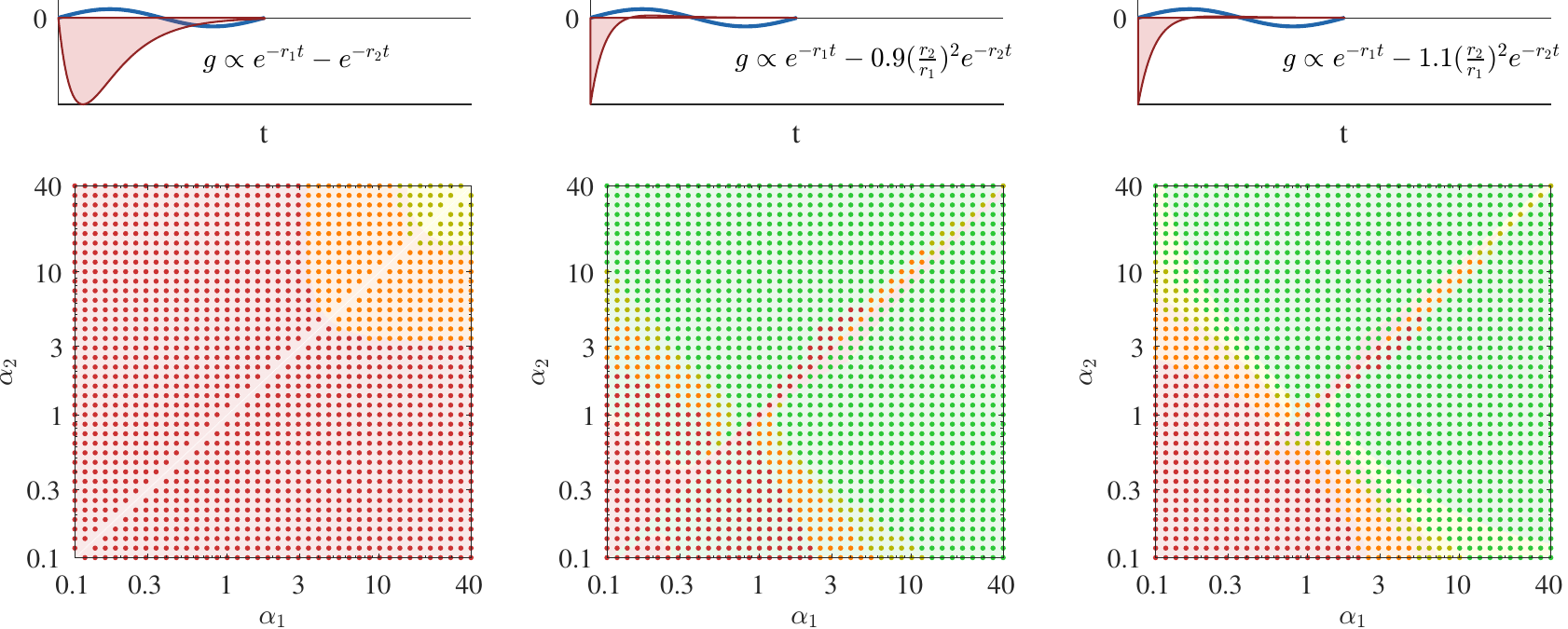}
    \caption{
    Stability phase diagram as in  Fig. \ref{fig:big} but for different response kernels $g(x,\Delta t)$.
    The middle and right columns add two exponentials, with slightly different choices of additive constant than that in Fig. \ref{fig:big} right. The left column shows two subtracted exponentials, equivalent to a convolution of two exponentials. The stability phase diagrams are analogous to Fig. \ref{fig:big} B. It is worth noting that the left column is qualitatively similar to a single exponential in that there is no stable region as $f\rightarrow1$ and that the left and right columns are no longer symmetric but do not deviate substantially from Fig. \ref{fig:big} B right. }
    \label{fig:SFIG}
\end{figure}

\noindent
In the main text, we compared three different response kernels. Here, we expand on that comparison and show an additional response kernel and how changing the weighting of the zero derivative kernel affects the phase behaviour. 
The additional response kernel we investigate is a convolution of two exponentials, 
\begin{equation}
    g(x,\Delta t)=\xi\omega_0^3\int_{-\infty}^\infty d t~ \theta( t)e^{-r_1(x) t}\theta(\Delta t-t)e^{-r_2(x)(\Delta t-t)}=\xi\omega_0^3\frac{1}{r_2(x)-r_1(x)}(e^{-r_1(x) \Delta t}-e^{-r_2(x)\Delta t})
\end{equation}
This convolution thus simplifies to a sum with a relative prefactor of $-1$. The left column of Fig. \ref{fig:SFIG} shows that this behaves in a way similar to an approximate derivative but without a region of true stability. 

The other changes tried were to the zero-derivative response kernel,
\begin{equation}
g(x,\Delta t) =\theta(\Delta t)\:\xi\omega_{0}^{2}\Big[e^{-r_1(x)\Delta t}-\frac{r_2(x)^2}{r_1(x)^2}e^{-r_2(x)\Delta t}\Big].
\label{eq:AIGkern}
\end{equation}
This kernel is motivated by having the property $\int_0^\infty d\Delta t g(x,\Delta t)\Delta t=0 \leftrightarrow \partial_\omega\tilde{g}(x,0)=0$. We found that it is stable for a large fraction of parameter space. To ensure this is not the product of fine-tuning, we change it so the derivative no longer integrates to exactly 0 in Fig \ref{fig:SFIG}. The phase diagram looks very similar except for some small changes close to the diagonal. The most important feature, stability for a large range of $\alpha_j$, remains true, indicating that our results do not require fine-tuning. 

\begin{figure}
    \centering
    \includegraphics[width=0.8\linewidth]{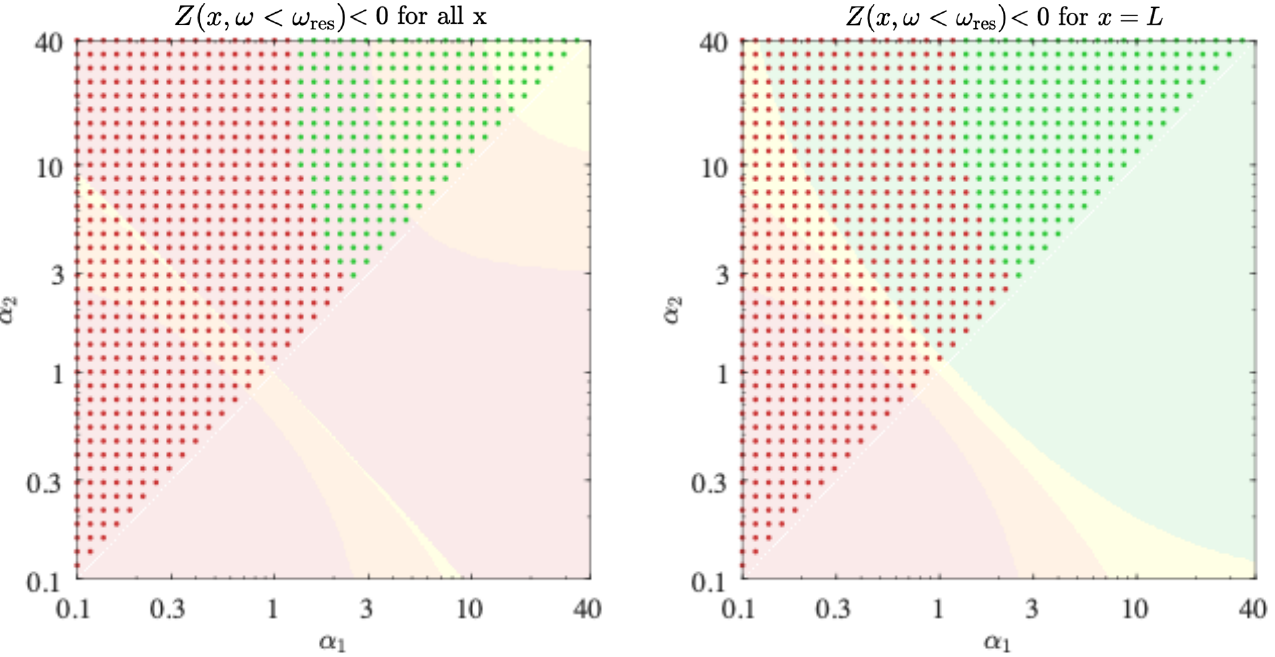}
    \caption{Combination of two response kernels from Eq. \ref{PAD0}. Phase diagrams are plotted in the same style as Fig. \ref{fig:big}B. In the left column, the analytic criteria we use for instability is $\relnetfriction(x,\omega)<0$ for $\omega<\omega_\text{res}(x)$ for all values of $x$, while for the right column we use $\relnetfriction(x,\omega)<0$ for $\omega<\omega_\text{res}$ only at $x=L$. Neither criterion fully predicts the simulation, unlike Fig. \ref{fig:big}B; however, examining friction at $x=L$ yields better predictions. 
    \label{fig:PA_D0}}
\end{figure}

The other change to response kernels we tried was having a concatenation of two different response kernel,
    \begin{align}
    \label{PAD0}
        g(x,\Delta t)&= \xi\omega_0^2(e^{-r_1(x)\Delta t}- (\frac{r_1(x)}{r_2(x)})e^{-r_2(x)\Delta t})~~~~~~x<\frac{L}{2}\\
        g(x,\Delta t)&= \xi\omega_0^2(e^{-r_1(x)\Delta t} -(\frac{r_1(x)}{r_2(x)})^2e^{-r_2(x)\Delta t})~~~~~~x\geq\frac{L}{2}
    \end{align}
The corresponding stability phase diagram is shown in Fig. \ref{fig:PA_D0}, and we can see from the results that what is most important is the behaviour for the later half of the BM. This is expected since the latter half of the BM has lower resonant frequencies and would have a larger contribution to the extended modes.   

\section{Effect of friction $\xi_\mathrm{ow}$ at the oval window}
\label{SI:OW}
\begin{figure}
    \centering
    \includegraphics[width=\linewidth]{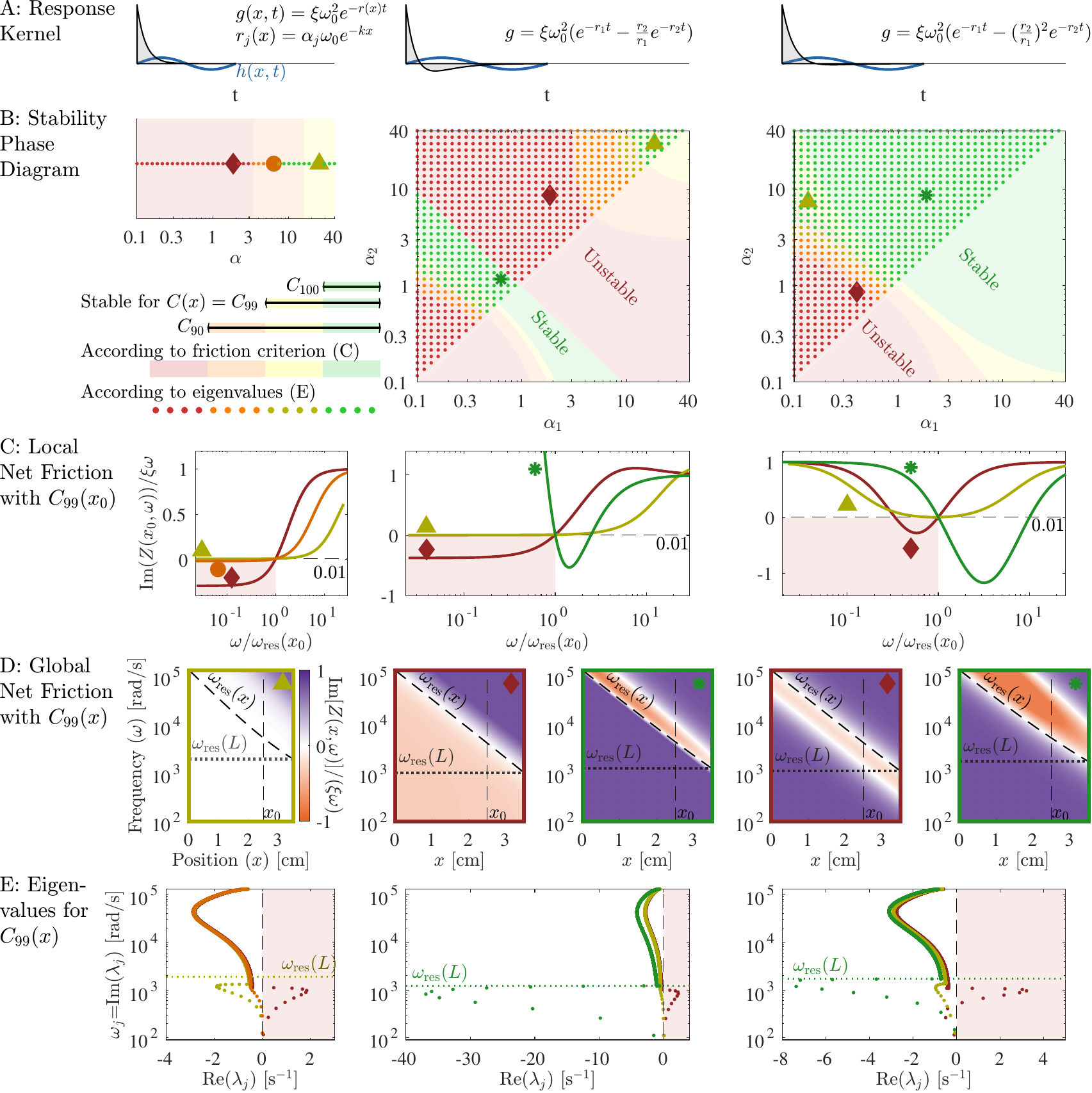}
    \caption{
    \textbf{The effect of increasing friction at the oval window.}
    This figure is identical to Fig.~3 in the main text except for using $\xi_\text{ow}=50000 \text{s}^{-1}$, a value 100 times greater than normal. Points marked by large symbols are chosen to be the same as those from the main text, even if their stability should indicate a different colour. 
    The criterion on $\relnetfriction(x,w)$ is completely unchanged (background shading in panel B), only the eigenvalues are different (dots in panel B).
    Note that at large values for $\alpha$, the simulations shows stable points even if friction at low values of $\omega$ is negative.}
    \label{fig:LF}
\end{figure}

\noindent
Two things can impact the stability of the extended modes. The first explained in the main text is the form of response kernel; more specifically, whether or not the response kernel causes $\Im(Z(x,\omega))<0$ for some values $\omega<\omega_\text{res}(x)$. The imaginary part of impedance determines if an extended mode gains energy as it travels along the cochlea. The other not discussed in main text is $\xi_\text{ow}$; this is the amount of energy lost at the left end of the cochlea. Fig. \ref{fig:LF} shows the behaviour of the three response kernels discussed in the main text when $\xi_\text{ow}$ is multiplied by 100. One can see that eigenvalues are stable for a larger set $\alpha_i$ than before. This effect is particularly noticeable for larger values of $\alpha_i$. It is important to note that Fig \ref{fig:LF} C,D are unchanged as the behaviour along the cochlea is unaffected by $\xi_\text{ow}$. It is also worth noting that in Fig \ref{fig:LF} E, we can see a large dip in the real part of localized modes that corresponds to a frequency close to that of $\omega_\text{ow}$. This is once again an effect of increasing $\xi_\text{ow}$.

\section{Self-tuning with coloured noise}\label{SI:colour}
\begin{figure}
    \centering
    \includegraphics[width=\linewidth]{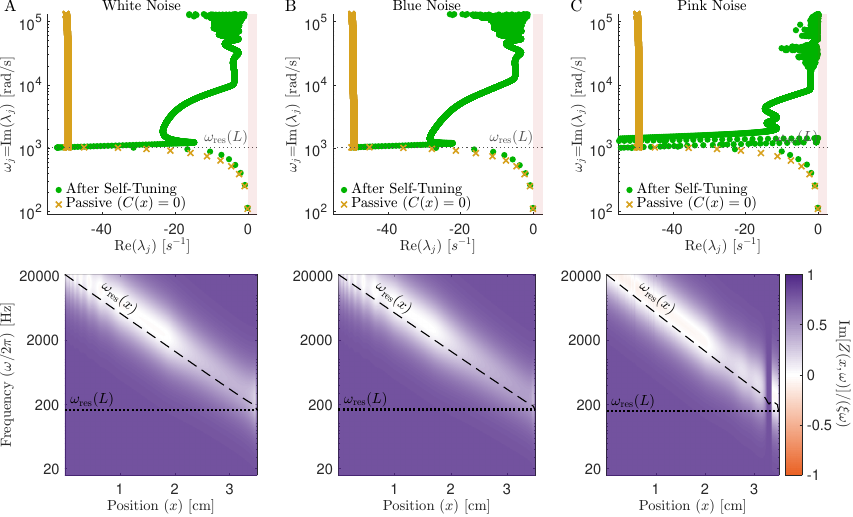}
    \caption{\textbf{Eigenvalue structure (top) and relative net friction $\relnetfriction$ (bottom) after self-tuning for different noise spectra}. This figure is similar to Fig.~\ref{fig:tune} in the main text, but for different choices of noise spectrum. Here, the target $h_0(x)$ has been set to be five times the passive RMS height (A) tuned with white noise, (B) tuned with blue noise, and (C) tuned with white noise. All systems reach a qualitatively similar equilibrium, with deviations for small frequency modes in the case of pink noise, likely due to boundary effects and interplay with extended modes.}
    \label{fig:tune-noise}
\end{figure}

\noindent
Figure~\ref{fig:tune-noise} is similar to Fig.~\ref{fig:tune} in the main text but for other choices of the noise spectrum. 
In particular, it compares the resulting eigenvalue structure and relative net friction for white noise, pink noise $\langle\eta(\omega)\eta(\omega)\rangle\propto \omega^{-1}$, and blue noise $\langle\eta(\omega)\eta(\omega)\rangle\propto \omega$.
Despite the changes in the noise spectrum, all systems exhibit qualitatively the same behavior: They reach a qualitatively similar equilibrium and all systems are amplified with reduced friction at resonance. The only noticeable difference occurs for the low-frequency modes in the case of pink noise, likely due to boundary effects and the interplay with extended modes.

\end{document}